\documentclass[12pt]{article}
\usepackage{amsmath,amssymb,mathrsfs,url,graphicx, color}
\usepackage{cite}

\title{Bohmian quantum gravity and cosmology}

\author{Nelson Pinto-Neto\footnote{COSMO - Centro Brasileiro de
Pesquisas F\'{\i}sicas -- CBPF, rua Xavier Sigaud, 150, Urca,
CEP22290-180, Rio de Janeiro, Brazil. E-mail: npintoneto@gmail.com} and  Ward Struyve\footnote{Mathematisches Institut, Ludwig-Maximilians-Universit\"at M\"unchen, Theresienstr.\ 39, 80333 M\"unchen, Germany. E-mail: ward.struyve@gmail.com}
}

\newcommand{\dd}{\mbox{d}}

\def\de{\delta}

\def\lam{\lambda}

\def\ka{\kappa}

\def\pa{\partial}

\def\th{\theta}
\def\al{\alpha}

\def\ka{\kappa}
\def\ii{\textrm i}
\def\ee{\textrm e}
\def\e{\textrm e}

\def\setR{\mathbb{R}}

\newcommand{\db}{de$\,$Broglie}

\newcommand{\dbb}{de$\,$Broglie-Bohm}

\newcommand{\be}{\begin{equation}}
\newcommand{\en}{\end{equation}}
\newcommand{\bi}{\begin{itemize}}
\newcommand{\ei}{\end{itemize}}

\begin{document}
\maketitle

\begin{abstract}
\noindent
Quantum gravity aims to describe gravity in quantum mechanical terms. How exactly this needs to be done remains an open question. Various proposals have been put on the table, such as canonical quantum gravity, loop quantum gravity, string theory, etc. These proposals often encounter technical and conceptual problems. In this chapter, we focus on canonical quantum gravity and discuss how many conceptual problems, such as the measurement problem and the problem of time, can be overcome by adopting a Bohmian point of view. In a Bohmian theory (also called pilot-wave theory or \dbb\ theory, after its originators \db\ and Bohm), a system is described by certain variables in space-time such as particles or fields or something else, whose dynamics depends on the wave function. In the context of quantum gravity, these variables are a space-time metric and suitable variable for the matter fields (e.g., particles or fields). In addition to solving the conceptual problems, the Bohmian approach yields new applications and predictions in quantum cosmology. These include space-time singularity resolution, new types of semi-classical approximations to quantum gravity, and approximations for quantum perturbations moving in a quantum background.
\end{abstract}

\section{Introduction}
Quantum theory arose as a description of the world on the smallest scales. It culminated in the standard model of high energy physics which describes the electro-weak and strong interaction. Our current best theory for gravity, on the other hand, is the general theory of relativity which is a classical theory about space-time and matter. While each theory is highly successful in its own domain, quantum theory on small scales and general relativity on large scales, it is as yet unknown how to harmonize both theories. Matter seems fundamentally quantum mechanical. So a theory of gravity should take this into account. One way to do this is by assuming gravity to be quantum as well. The most conservative approach to achieve this, is by applying the usual quantization techniques, which turn classical theories into quantum theories, to general relativity. This results in a theory called canonical quantum gravity, described by the Wheeler-DeWitt equation. While these quantization techniques have proved to be enormously successful in the context of the standard model, there is no guarantee for success in the gravitational case. It is therefore crucial to look for possible experimental tests. However, it has been hard to extract possible predictions from canonical quantum gravity because the theory is problematic, not only due to the many technical issues surrounding the Wheeler-DeWitt equation, such as dealing with its infinities, but also due to the conceptual problems, such as the measurement problem and the problem of time, when it is considered in the framework of orthodox quantum theory.

In this chapter we show that a significant progress can be made by considering the canonical quantization of gravity from the Bohmian point of view. In a Bohmian theory, a system is described by certain variables in space-time such as particles or fields or something else, whose dynamics depends on the wave function \cite{bohm93,holland93b,duerr09,duerr12}. In the context of non-relativistic Bohmian mechanics, these variables are particle positions. So in this case there are actual particles whose motion depends on the wave function. Bohmian mechanics can also be extended to quantum field theory \cite{struyve11a}, where the variables may be particles or fields, and to canonical quantum gravity \cite{vink92,shtanov96,goldstein04,pinto-neto05a,pinto-neto13}. In the context of quantum gravity, the extra variables are a space-time metric and whatever suitable variable for the matter fields. This Bohmian formulation of quantum gravity solves the aforementioned conceptual problems. As such it can make unambiguous predictions in, for example, quantum cosmology. The aim of this chapter is to give an introduction to Bohmian quantum gravity, explain how it solves the conceptual problems with the conventional approach, and give examples of practical applications and novel predictions.

The chapter is organized as follows. We start with an introduction to non-relativistic Bohmian mechanics in section \ref{nrbm}, and highlight some important properties that will also be used in the context of quantum gravity. In section \ref{cqg}, we introduce canonical quantum gravity and discuss the conceptual problems that appear when trying to interpret the theory in the context of orthodox quantum theory. In section \ref{bcqg}, we turn to Bohmian quantum gravity and explain how it solves these conceptual problems. In section \ref{mini-superspace}, we discuss mini-superspace models. These are simplified models of quantum gravity, which assume certain symmetries such as homogeneity and isotropy. In the context of such models, we consider the problem of space-time singularities in section \ref{singularities}. In section \ref{perturbations}, the mini-superspace models are extended to include perturbations. These perturbations are important in the description of structure formation. Bohmian approximation techniques will be employed to obtain tractable equations of motion. Observational consequences will be discussed for a particular model with Bohmian matter bounces. In addition, we show how the problem of the quantum-to-classical limit in inflationary and bouncing models is solved. Finally, in section \ref{scg}, we discuss a new approach to semi-classical gravity, which treats gravity classically and matter quantum mechanically, based on Bohmian mechanics.

\section{Non-relativistic Bohmian mechanics}\label{nrbm}
Non-relativistic Bohmian mechanics is a theory about point-particles in physical space moving under the influence of the wave function \cite{bohm93,holland93b,duerr09,duerr12}. The equation of motion for the configuration $X=({\bf X}_1,\dots,{\bf X}_n)$ of the particles, called the {\em guidance equation}, is given by{\footnote{Throughout the paper we assume units in which $\hbar=c=1$.}}
\be
{\dot X}(t) = v^\psi(X(t),t) \,,
\label{0.01}
\en
where $v^\psi=({\bf v}^\psi_1, \dots , {\bf v}^\psi_n)$, with
\be
{\bf v}^\psi_k = \frac{1}{m_k} {\textrm{Im}}\left( \frac{\boldsymbol{\nabla}_k \psi}{\psi} \right) =  \frac{1}{m_k} {\boldsymbol \nabla}_k S
\label{0.02}
\en
and $\psi = |\psi| \ee^{\ii S}$. The wave function $\psi(x,t)=\psi({\bf x}_1,\dots,{\bf x}_n)$ itself satisfies the non-relativistic Schr\"odinger equation
\be
\ii \pa_t \psi(x,t) = \left( - \sum^n_{k=1} \frac{1}{2m_k} \nabla^2_k + V(x) \right) \psi(x,t) \,.
\label{0.03}
\en

For an ensemble of systems all with the same wave function $\psi$, there is a distinguished distribution given by $|\psi|^2$, which is called the {\em quantum equilibrium distribution}. This distribution is {\em equivariant}. That is, it is preserved by the particle dynamics \eqref{0.01} in the sense that if the particle distribution is given by $|\psi(x,t_0)|^2$ at some time $t_0$, then it is given by $|\psi(x,t)|^2$ at all times $t$. This follows from the fact that any distribution $\rho$ that is transported by the particle motion satisfies the continuity equation
\be
\pa_t \rho + \sum^n_{k=1} {\boldsymbol \nabla}_k \cdot ({\bf v}^\psi_k \rho) = 0
\label{0.04}
\en
and that $|\psi|^2$ satisfies the same equation, i.e.,
\be
\pa_t |\psi|^2 + \sum^n_{k=1} {\boldsymbol \nabla}_k \cdot ({\bf v}^\psi_k |\psi|^2) = 0 \,,
\label{0.041}
\en
as a consequence of the Schr\"odinger equation.

It can be shown that for a typical initial configuration of the universe, the (empirical) particle distribution for an actual ensemble of subsystems within the universe will be given by the quantum equilibrium distribution \cite{duerr92a,duerr09,duerr12}. Therefore, for such a configuration Bohmian mechanics reproduces the usual quantum predictions.

Non-equilibrium distributions would lead to a deviation of the Born rule. While such distributions are atypical, they remain a logical possibility \cite{valentini91a}. However, it remains to be seen whether they are physically relevant.

Note that the velocity field is of the form $j^\psi/|\psi|^2$, where $j^\psi=({\bf j}^\psi_1,\dots,{\bf j}^\psi_n)$ with ${\bf j}^\psi_k=  {\textrm{Im}}( \psi^* \boldsymbol{\nabla}_k \psi )/m_k $ is the usual quantum current. In other quantum theories, such as for example quantum field theories and canonical quantum gravity, the velocity can be defined in a similar way by dividing the appropriate current by the density. In this way equivariance of the density will be ensured. (See \cite{struyve09a} for a treatment of arbitrary Hamiltonians.)

One motivation to consider Bohmian mechanics is the measurement problem. Orthodox quantum mechanics works fine for practical purposes. However, the measurement problem implies that orthodox quantum mechanics can not be regarded as a fundamental theory of nature. The problem arises from the fact that the wave function has two possible time evolutions. On the one hand there is the Schr\"odinger evolution, on the other hand there is wave function collapse. But it is unclear when exactly the collapse takes place. The standard statement is that collapse happens upon measurement. But which physical processes count as measurements? Which systems count as measurement devices? Only humans? Or rather humans with a PhD.\ \cite{bell90}? Bohmian mechanics solves this problem. In Bohmian mechanics the wave function never collapses; it always evolves according to the Schr\"odinger equation. There is no special role for measurement devices or observers. They are treated just as other physical systems.

There are two aspects of the theory whose analogue in the context of quantum gravity will play an important role. Firstly, Bohmian mechanics allows for an unambiguous analysis of the classical limit. Namely, the classical limit is obtained whenever the particles (or at least the relevant macroscopic variables, such as the center of mass) move classically, i.e., satisfy Newton's equation. By taking the time derivative of \eqref{0.01}, it is found that
\be
m_k {\ddot {\bf X}}_k(t) = -{\boldsymbol{\nabla}}_k (V(x)+Q^\psi(x,t))\big|_{x=X(t)}\,,
\label{0.07}
\en
where
\be
Q^\psi = -\sum^n_{k=1}\frac{1}{2m_k}\frac{\nabla^2_k |\psi|}{|\psi|}
\label{0.08}
\en
is the quantum potential. Hence, if the quantum force $-{\boldsymbol{\nabla}}_kQ^\psi$ is negligible compared to the classical force $-{\boldsymbol{\nabla}}_kV$, then the $k$-th particle approximately moves along a classical trajectory.

Another aspect of the theory is that it allows for a simple and natural definition for the wave function of a subsystem \cite{duerr92a,duerr09}. Namely, consider a system with wave function $\psi(x,y)$ where $x$ is the configuration variable of the subsystem and $y$ is the configuration variable of its environment. The actual configuration is $(X,Y)$, where $X$ is the configuration of the subsystem and $Y$ is the configuration of the other particles. The wave function of the subsystem $\chi(x,t)$, called the {\em conditional wave function}, is then defined as
\be
\chi(x,t) = \psi(x,Y(t),t).
\label{0.05}
\en
This is a natural definition since the trajectory $X(t)$ of the subsystem satisfies
\be
{\dot X}(t) = v^\psi(X(t),Y(t),t) = v^\chi(X(t),t) \,.
\label{0.06}
\en
That is, for the evolution of the subsystem's configuration we can either consider the conditional wave function or the total wave function (keeping the initial positions fixed). (The conditional wave function is also the wave function that would be found by a natural operationalist method for defining the wave function of a quantum mechanical subsystem \cite{norsen14}.) The time evolution of the conditional wave function is completely determined by the time evolution of $\psi$ and that of $Y$. The conditional wave function does not necessarily satisfy a Schr\"odinger equation, although in many cases it does. This wave function collapses during measurement situations. This explains the success of the collapse postulate in orthodox quantum mechanics. In the context of quantum gravity, the conditional wave function will be used to derive an effective time-dependent wave equation for a subsystem of the universe from a time-independent universal wave function.

\section{Canonical quantum gravity}\label{cqg}
Canonical quantum gravity is the most conservative approach to quantum gravity. It is obtained by applying the usual quantization techniques, which were so successful in high energy physics, to Einstein's theory of general relativity. The quantization starts with passing from the Lagrangian to the Hamiltonian picture and then mapping Poisson brackets to commutation relations of operators. Let us start with an outline of this procedure.

In general relativity, gravity is described by a Lorentzian space-time metric $g_{\mu \nu}(x)$, which satisfies the Einstein field equations
\be
G_{\mu \nu } = 8\pi G T_{\mu \nu},
\label{cqg1}
\en
where $G$ is the gravitational constant, $G_{\mu \nu }$ the Einstein tensor and $T_{\mu \nu}$ the energy-momentum tensor, whose form is determined by the type of matter. In order to pass to the Hamiltonian picture, a splitting of space and time is necessary. This is done by assuming a foliation of space-time into space-like hypersurfaces so that ${\mathcal M}$ is diffeomorphic to ${\mathbb R} \times \Sigma$, with $\Sigma$ a 3-surface. Coordinates $x^\mu=(t,{\bf x})$ can be chosen such that the time coordinate $t$ labels the leaves of the foliation and ${\bf x}$ are coordinates on $\Sigma$. In terms of these coordinates the space-time metric and its inverse can be written as
\be
g_{\mu \nu}=
\begin{pmatrix}
N^2 - N_i N^i & -N_i \\
-N_i & - h_{ij}
\end{pmatrix} \,,
\qquad
g^{\mu \nu}=
\begin{pmatrix}
\frac{1}{N^2} & \frac{-N^i}{N^2} \\
\frac{-N^i}{N^2} &   \frac{N^iN^j}{N^2}- h^{ij}
\end{pmatrix} \,,
\label{cqg2}
\en
where $N > 0$ is the lapse function, $N_i = h_{ij}N^j$ are the shift functions, and $h_{ij}$ is the induced Riemannian metric on the leaves of the foliation.

The geometrical meaning of the lapse and shift is the following \cite{misner73}. The unit vector field normal to the leaves is $n^\mu=(1/N,-N^i/N)$. The lapse $N(t,{\bf x})$ is the rate of change with respect to coordinate time $t$ of the proper time of an observer with four-velocity $n^{\mu}(t,{\bf x})$ at the point $(t,{\bf x})$. The lapse function also determines the foliation. Lapse functions that differ only by a factor $f(t)$ determine the same foliation. Lapse functions that differ by more than a factor $f(t)$ determine different foliations. $N^i(t,{\bf x})$ is the rate of change with respect to coordinate time $t$ of the shift of the points with the same coordinates ${\bf x}$ when we go from one hypersurface to another. Different choices of $N^i$ correspond to different choices of coordinates on the space-like hypersurfaces.

The Hamiltonian picture makes it manifest that the functions $N$ and $N^i$ are arbitrary functions of space-time. The spatial metric $h_{ij}$ satisfies non-trivial dynamics, corresponding to how it changes along the succession of space-like hypersurfaces. The arbitrariness of $N$ and $N^i$ arises from the space-time diffeomorphism invariance of the theory (i.e., the invariance under space-time coordinate transformations). The motion of $h_{ij}$ does not depend on the foliation. That is, the evolution of an initial 3-metric on a certain space-like hypersurface to a future space-like hypersurface does not depend on the choice of intermediate hypersurfaces. Spatial metrics $h_{ij}$ that differ only by spatial diffeormophisms determine the same physical 3-geometry. The dynamics is therefore called {\em geometrodynamics}. A succession of 3-metrics determines a 4-geometry.

Canonical quantization introduces an operator ${\widehat h}_{ij}({\bf x})$ which acts on wave functionals $\Psi(h_{ij})$, which are functionals of metrics on $\Sigma$. In the presence of matter, the wave functional also depends on the matter degrees of freedom. But we assume just gravity for now. The wave functional satisfies the functional Schr\"odinger equation
\be
\ii \pa_t \Psi = \int d^3 x \left(N {\widehat {\mathcal H}} + N^i {\widehat {\mathcal H}}_i  \right) \Psi\,,
\label{cqg10}
\en
where
\be
{\mathcal H}  =  - 16\pi G G_{ijkl}\frac{\delta}{\delta h_{ij}} \frac{\delta}{\delta h_{kl}} + \frac{h^{1/2}}{16\pi G}( 2\Lambda - R^{(3)})\,, \qquad {\mathcal H}_i  =  -2 h_{il}D_j\frac{\delta }{\delta h_{jl}} \,,
\label{cqg11}
\en
with $G_{ijkl}$ the DeWitt metric (which depends on the 3-metric), $h$ the determinant of $h_{ij}$, $R^{(3)}$ the 3-curvature and $\Lambda$ the cosmological constant. In addition to the Schr\"odinger equation \eqref{cqg10}, the wave functional has to satisfies the constraints:
\be
{\mathcal H} \Psi = 0 \,,
\label{cqg12}
\en
\be
{\mathcal H}_i \Psi = 0\,, \quad i=1,2,3 \,.
\label{cqg13}
\en
This immediately implies that
\be
\pa_t \Psi = 0 \,,
\en
i.e., the wave function does not depend on time. This is the source of the problem of time. The constraint \eqref{cqg13}, which is called the {\em diffeomorphism constraint}, implies invariance of $\Psi$ under infinitesimal spatial coordinate transformations. The constraint \eqref{cqg12} is called the Wheeler-DeWitt equation and is believed to somehow contain the time-evolution.

There are technical problems with this theory. Namely the functional Schr\"odinger equation is merely formal and needs to be regularized. In addition, a proper Hilbert space needs to be found. The theory is also not renormalizable. The latter does not necessarily mean a failure of the theory, but merely that the usual perturbation techniques do not work.

In addition to the technical problems, there are also conceptual problems. First of all, there is the measurement problem, which carries over from non-relativistic quantum mechanics. However, in this case the problem is even more severe. Namely, the aim is to describe the whole universe (albeit with simplified models), and then there are no outside observers or measurement devices that could collapse the wave function. In addition, the aim is also to describe, for example, the early universe, and there are no observers or measurement devices present even within the universe. Second, there is the problem of time \cite{isham92,kuchar92,kiefer04}. The wave function is static. So how can time evolution be explained in terms of such a wave function? How can we tell from the theory whether the universe is expanding or contracting, or running into a singularity? Finally, there is the problem of what it means to have a space-time singularity. The universe is described solely by a wave function, but there is no actual metric. Various definitions of what a singularity could mean have been explored \cite{dewitt67a,ashtekar06b,ashtekar06c,ashtekar06d,ashtekar08,ashtekar11}: that the wave function has support on singular metrics, that the wave function is peaked around singular metrics, that the expectation value of the metric operator is singular, etc. Although these definitions may have something to say about the occurrence or non-occurrence of singularities, neither of these is completely satisfactory. In fact, since there is merely the wave function, one might even consider the question about space-time singularities as off-target, since it is the dynamics of the wave function that needs to be well-defined (which in this case amounts to finding solutions to the constraints). The question of the meaning of singularities is important because loop quantum gravity is believed to eliminate singularities, while canonical quantum gravity is not.

\section{Bohmian canonical quantum gravity}\label{bcqg}
In the Bohmian approach to canonical quantum gravity, there is an actual 3-metric $h_{ij}$, whose motion is given by
\be
{\dot{h}}_{ij} = 32 \pi G NG_{ijkl}\frac{\delta S}{\delta h_{kl}} + D _i N_j + D _j N_i,
\label{bcqg1}
\en
with $\Psi=|\Psi|\ee^{\ii S}$ and $D_i$ the 3-dimensional covariant derivative. This equation can be obtained by considering a suitable current as explained in section \ref{nrbm} \cite{goldstein04}. It can also be obtained by considering the classical Hamilton equation and replacing the momentum conjugate to $h_{ij}$ by $\delta S/\delta h_{kl}$ \cite{pinto-neto05a}. So, just as in the case of general relativity the theory concerns geometrodynamics, i.e., it is about an evolving 3-geometry.

Different choices of the shift vectors $N_i$ correspond to different coordinates on the spatial hypersurfaces. The Bohmian dynamics does not depend on the choice of coordinates on the spatial hypersurfaces. That is, the dynamics is invariant under spatial diffeomorphisms. A convenient choice is to take the $N_i = 0$ \cite{goldstein04}. The lapse function $N>0$ determines the foliation. Lapse functions that differ only by a factor $f(t)$ (which only depends on the time $t$ which labels the leaves of the foliation) determine the same foliation. The Bohmian dynamics does not depend on such different choices. Such a difference merely corresponds to a time-reparameterization. However, different lapse functions that differ by more than a factor $f(t)$ generically yield different Bohmian dynamics \cite{shtanov96,pinto-neto99,goldstein04,shojai04}. That is, if we consider the motion of an initial 3-metric along a certain space-like hypersurface and let it evolve according to the dynamics given in Eq.~(\ref{bcqg1}) to a future space-like hypersurface, then the final 3-metric will depend on the choice of lapse function or, in other words, on the choice of intermediate hypersurfaces. This was shown in detail in \cite{pinto-neto99}. This is unlike general relativity, where there is foliation-independence. So in the Bohmian theory a particular choice of lapse function or foliation needs to be made. As such the theory is not generally covariant. This is akin to the situation in special relativity where the non-locality (which is unavoidable for any empirically adequate quantum theory, due to Bell's theorem) is hard to combine with Lorentz invariance. In that context, it is simpler to assume a preferred reference frame or foliation. The extent to which this extra space-time structure can be eliminated and the theory be made fully Lorentz invariant is discussed in \cite{duerr14}. One possibility is to let the foliation be determined in a covariant way by the wave function. Perhaps a similar approach can be taken in the case of quantum gravity \cite{goldstein04}, but so far no concrete examples have been considered.

This theory solves the aforementioned conceptual problems with interpreting canonical quantum gravity. First of all there is no measurement problem. Measurement devices or observers do not play a fundamental role in the theory. Second, even though the wave function is static, the evolution of the 3-metric is generically time-dependent. This evolution will, for example, indicate whether the universe is expanding or contracting.  Effective time-dependent Schr\"odinger equations can be derived for subsystems by considering the conditional wave function. Suppose, for example, that we are dealing with a scalar field in the presence of gravity, for which the wave function is $\Psi(h_{ij},\varphi)$. Then for a solution $(h_{ij}({\bf x},t),\varphi({\bf x},t))$ of the guidance equations, one can consider the conditional wave function for the scalar field: $\chi(\varphi,t) = \Psi(h_{ij}({\bf x},t),\varphi)$. In certain cases $\chi$ will approximately satisfy a time-dependent Schr\"odinger equation. Explicit examples will be given in sections \ref{perturbations} and \ref{scg}. There is a similar procedure to derive the time-dependent Schr\"odinger equation in the context of orthodox quantum theory \cite{kiefer04}. In this procedure, a classical trajectory $h_{ij}({\bf x},t)$ is plugged into the wave function, rather than a Bohmian one. While this procedure indeed gives a time-dependent wave function for the scalar field, it seems rather ad hoc. In Bohmian mechanics, the conditional wave function is motivated by the fact that the velocity, and hence the evolution of the actual scalar field $\varphi$, can be expressed in terms of either the conditional or the universal wave function, cf.\ Eq.\ \eqref{0.06}. The same trajectory is obtained. In the context of orthodox quantum mechanics there seems to be no justification for conditionalizing the wave function on a classical trajectory. Moreover, Bohmian mechanics is broader in its scope because it not only allows to conditionalize on classical paths but also on non-classical ones. This allows to go beyond the usual semi-classical analysis, as will be shown in section~\ref{perturbations}.

Finally, the meaning of space-time singularities becomes unambiguous. Namely, it is the same meaning as in general relativity: there is a singularity whenever the actual metric becomes singular. We will discuss examples in section \ref{singularities}.

Taking the time derivative of the guidance equation \eqref{bcqg1} and using the expression \eqref{cqg2} for the metric $g_{\mu \nu}$, the modified Einstein equations are obtained:
\be
G_{\mu \nu } = 8\pi G T_{Q\mu \nu} ,
\label{modifiedee}
\en
where $T_{Q}^{\mu \nu}$ is an energy-momentum tensor of purely quantum mechanical origin (there is no matter in this case). It is given by
\be T_{Q}^{\mu \nu}(x) = - \frac{2}{\sqrt{-g(x)}} \frac{\de}{\de g_{\mu \nu}(x)} \int d^4y N(y)Q(y), \en
with
\be Q = -16\pi G G_{ijkl} \frac{1}{|\Psi|}  \frac{\delta^2 |\Psi|}{\delta h_{ij}\delta h_{kl}} \en
the quantum potential. More explicitly:
\be T_{Q}^{00}({\bf x},t) = \frac{1}{N^2({\bf x},t) } \frac{Q({\bf x},t)}{\sqrt{h({\bf x},t)}}    , \quad  T_{Q}^{0i}({\bf x},t) = T_{Q}^{i0}({\bf x},t) =  - \frac{N^i({\bf x},t)}{N^2 ({\bf x},t)} \frac{Q({\bf x},t)}{\sqrt{h({\bf x},t)}}, \en 
\begin{multline}
T_{Q}^{ij}({\bf x},t) = \left(\frac{N^i({\bf x},t)N^j({\bf x},t)}{N^2({\bf x},t)} - h^{ij}({\bf x},t)\right)  \frac{Q({\bf x},t)}{\sqrt{h({\bf x},t)}}  \\
- \frac{2}{N({\bf x},t)\sqrt{h({\bf x},t)}}  \int d^3y  N({\bf y},t)\sqrt{h({\bf y},t)} \frac{\de }{\de h_{ij}({\bf x})} \left( \frac{Q({\bf y},t)}{\sqrt{h({\bf y},t)}} \right)  .
\end{multline}
While Eq.~\eqref{modifiedee} was written in covariant form, it is not generally covariant due to the preferred choice of lapse function. When $T_{Q\mu \nu}$ vanishes, we obtain the classical equations which are generally covariant.

The classical limit is obtained whenever $T_{Q\mu \nu}$ is negligible. The deviation from classicality also often causes singularities to be avoided, as we will see in the next section.

Note that if $T_{Q\mu \nu} \sim g_{\mu \nu}$, then it would act as a cosmological constant. It is interesting to speculate whether the observed cosmological constant can indeed be of quantum origin \cite{squires92,pinto-neto03}.

In orthodox quantum mechanics, it is important to consider a particular Hilbert space (which is difficult in this case). However, from the Bohmian point of view, this is not necessary. It is necessary however that the Bohmian dynamics be well-defined. For example, in the context of non-relativistic Bohmian mechanics, a plane wave is not in the Hilbert space but the corresponding trajectories are well-defined. They are just straight lines.

What about probabilities? The Bohmian dynamics preserves the density $|\Psi(h_{ij})|^2$. However, this density is not normalizable (with respect to some appropriate measure ${\mathcal D}h$){\footnote{This is rather formal and requires some mathematical rigor to make precise, but similar statements can be made in the context of mini-superspace models, to be discussed in the next section, which are mathematically precise.}} due to the constraints, so it can not immediately be used to make statistical predictions. For certain predictions, it is not required, because we only have a single universe. On the other hand, statistical predictions play an important role in case where one can identify subsystems within the universe. We will see later on how this can be accomplished.

\section{Mini-superspace}\label{mini-superspace}
The Wheeler-DeWitt equation (\ref{cqg12}) and the diffeomorphism constraint \eqref{cqg13} are very complicated functional differential equations which are hard to solve. In order to make the equations tractable, one often assumes certain symmetries like translation and rotation invariance. This reduces the number of degrees of freedom to a finite one. It is physically justified because we are interested in applying this formalism to the primordial universe, and observations indicate that it was very homogeneous and isotropic at these early times. Even today, at large scales, the universe seems to be spatially homogeneous and isotropic.

Rather than deriving the quantum mini-superspace models from the full quantum theory, they are obtained from the canonical quantization of the reduced classical theory, using the action obtained from the full action upon imposition of the considered symmetries. It is as yet unclear to what extent these reduced quantum theories follow from the full quantum theory. The starting point is to express the metric $h_{ij}$ and the matter degrees of freedom in terms of a finite number of variables, say $q^a$, $a=1,\dots,n$. By moving from the Lagrangian to the Hamiltonian picture one obtains a Hamiltonian which generally has the form
\be
H = N {\mathcal H} = N \left( \frac{1}{2} f^{ab}(q) p_a p_b + U (q)\right),
\en
where $p_a$ are the momenta conjugate to $q^a$, $f^{ab}(q)$ is a symmetric function of the $q$'s whose inverse plays the role of a metric on $q$-space and $U$ is a potential. $N(t)>0$ is the lapse function, which is arbitrary. It does not depend on space, since we can choose a foliation where the fields are homogeneous. The dynamics is generated by $H$, but constrained to satisfy ${\mathcal H}=0$. This yields the equations of motion
\begin{align}
\dot{q}^a &= N f^{ab}p_b, \\
\dot{p}_a &= -N \left(\frac{1}{2} \frac{\pa f^{bc}(q)}{\pa q^a} p_b p_c + \frac{\pa U(q)}{\pa q^a} \right) ,
\label{minicl}
\end{align}
together with the constraint
\be
\frac{1}{2} f^{ab}(q) p_a p_b + U (q) =0.
\en
Since the lapse is arbitrary, the dynamics is time-reparameterization invariant. The time parameter $t$ is itself unobservable. Physical clocks should be modeled in terms of one of the variables $q$, say $q^a$. Namely, if $q^a$ changes monotonically with $t$, it can be treated as a clock variable, and $t$ could be eliminated by inverting $q^a(t)$.

Quantization of this model is done by introducing an operator
\be
{\widehat {\mathcal H}} = - \frac{1}{2} \frac{1}{\sqrt{f}}\frac{\pa}{\pa q^a} \left( f^{ab}(q) \sqrt{f} \frac{\pa}{\pa q^b} \right)  + U (q),
\en
where $f$ is the determinant of the inverse of $f^{ab}$, and which acts on wave functions $\psi(q)$. Here, interpreting $f^{ab}$ as a metric on $q$-space, the Laplace-Beltrami operator was chosen, as is usually done \cite{halliwell86}. This imposes an operator ordering choice. The Wheeler-DeWitt equation now reads
\be
{\widehat {\mathcal H}} \psi = 0
\en
and the guidance equations are
\be
{\dot q}^a = N f^{ab}\frac{\pa S}{\pa q^b} ,
\label{gems}
\en
where $\psi = |\psi| \ee^{\ii S}$. The function $N(t)$ is again the lapse function. It is arbitrary, just as in the classical case, which implies that the dynamics is time-reparameterization invariant.

The continuity equation implied by the Wheeler-DeWitt equation is
\be
\frac{\pa}{\pa q^a} \left(f^{ab}\frac{\pa S}{\pa q^b}  |\psi|^2 \right) = 0,
\en
which implies that the density $|\psi|^2$ is preserved by the Bohmian dynamics. As mentioned in section \ref{nrbm}, this motivates the choice of the guidance equations \eqref{gems}.

Denoting $p_a = \pa S/\pa q^a$, the Bohmian dynamics implies the classical equations \eqref{minicl}, but with the potential $U$ replaced by $U+Q$, with
\be
Q = - \frac{1}{2\sqrt{f} |\psi|} \frac{\pa}{\pa q^a} \left( f^{ab}\sqrt{f}\frac{\pa}{\pa q^b} |\psi| \right)
\en
the quantum potential.

In the next section, when considering the question of space-time singularities, we will consider two types of mini-superspace models. A Friedmann-Lema\^itre-Robertson-Walker metric coupled to respectively a canonical scalar field and a perfect fluid.

\section{Space-time singularities}\label{singularities}
According to general relativity, space-time singularities such as a big bang or big crunch are generically unavoidable. This is usually taken as signaling the limited validity of the theory and the hope is that a quantum theory for gravity will eliminate the singularities. In Bohmian quantum gravity, we can unambiguously analyse the question of singularities because there is an actual metric and the meaning of singularities is the same as in general relativity.

In this section, the question of big bang or big crunch singularities is considered in the simple case of a homogeneous and isotropic metric respectively coupled to a homogeneous scalar field (with zero matter potential \cite{pinto-neto12b,struyve17b} and with exponential matter potential \cite{colin17,bacalhau17}) and to a perfect fluid, modelled also by a scalar field \cite{acacio98,alvarenga02,pinto-neto05b,pinto-neto13}. After considering the Wheeler-DeWitt quantization, we also consider the loop quantization of the former model \cite{struyve17b}. In the Wheeler-DeWitt case, there may be singularities depending on the wave function and the initial conditions. In the case of loop quantization there are no singularities.

Anisotropic models are discussed in \cite{pinto-neto13}.

\subsection{Mini-superspace - canonical scalar field}\label{mcsf}
The simplest example of a mini-superspace model is that of a homogeneous and isotropic Friedmann-Lema\^itre-Robertson-Walker (FLRW) metric coupled to a homogeneous scalar field. The metric is
\be
\dd s^2 = N(t)^2 \dd t^2 - a(t)^2  \dd \Omega^2_k ,
\label{s1}
\en
where $N$ is the lapse function, $a=\ee^{\al}$ is the scale factor, and $\dd \Omega^2_k$ is the spatial line-element on three-space with constant curvature $k$. In the classical theory, the coupling to a homogeneous scalar field $\varphi$ is described by the Lagrangian
\be
L = N \ee^{3\al} \left( \ka^2 \frac{\dot \varphi^2}{2N^2} - \ka^2 V_M  - \frac{\dot \al^2}{2N^2} - V_G\right),
\label{s2}
\en
where $\kappa = \sqrt{4\pi G/3}$, with $G$ the gravitational constant, $V_M$ is the potential for the scalar field, $V_G =- \frac{1}{2}k\ee^{-2\al} + \frac{1}{6}\Lambda$, and $\Lambda$ is the cosmological constant \cite{halliwell91b,struyve15}. The classical equations of motion are
\be
\frac{d}{dt}\left( \frac{\ee^{3\al} \dot \varphi}{N}\right) + N \ee^{3\al} \pa_\varphi V_M =0,
\label{s3}
\en
\be
 \frac{\dot \al^2}{N^2} = 2 \ka^2 \left( \frac{\dot \varphi^2}{2N^2} + V_M\right) + 2 V_G.
 \label{s4}
 \en
The latter equation is the Friedmann equation. The acceleration equation, which corresponds to the second-order equation for $\al$, follows from \eqref{s3} and \eqref{s4}.

There is a big bang or big crunch singularity when $a=0$, i.e., $\al \to -\infty$. This singularity is obtained for generic solutions, as was shown by the Penrose-Hawking theorems.

Canonical quantization of the classical theory leads to the Wheeler-DeWitt equation
\be
\left[- \frac{1}{2\ee^{3\al}} \pa^2_\varphi + \frac{\ka^2}{2\ee^{3\al}}\pa^2_\al + \ee^{3\al}\left(V_M + \frac{1}{\ka^2} V_G\right) \right]\psi(\varphi,\al) =0.
\label{s5}
\en
In the Bohmian theory \cite{vink92,pinto-neto12b} there is an actual scalar field $\varphi$ and an actual FLRW metric of the form \eqref{s1}, whose time evolution is determined by
\be
\dot \varphi = \frac{N}{e^{3\alpha}} \pa_\varphi S , \quad \dot \al = - \frac{N}{e^{3\alpha}} \kappa^2 \pa_\al S .
\label{s6}
\en
It follows from these equations that
\be
\frac{d}{dt}\left( \frac{\ee^{3\al} \dot \varphi}{N}\right) + N \ee^{3\al} \pa_\varphi (V_M + Q_M)=0,
\en
\be
\frac{\dot \al^2}{N^2} = 2 \ka^2 \left( \frac{\dot \varphi^2}{2N^2} +(V_M + Q_M) \right) + 2 (V_G + Q_G),
\label{s6.01}
\en
where
\be
Q_M = -\frac{1}{2\ee^{6\al}} \frac{\pa^2_\varphi |\psi|}{|\psi|} , \qquad Q_G = \frac{\kappa^4}{2\ee^{6\al}} \frac{\pa^2_\al |\psi|}{|\psi|}
\en
are respectively the matter and the gravitational quantum potential. These equations differ from the classical ones by the quantum potentials.

In order to discuss the singularities, we consider the case of a free massless scalar field and that of an exponential potential.

\subsubsection{Free massless scalar field}
In the case of $V_M = V_G = 0$, the classical equations lead to
\be
\dot \varphi =  \frac{N}{\ee^{ 3\al}} c , \qquad \dot \al = \pm  \frac{N}{\ee^{ 3\al}} \kappa^2 c ,
\label{s7}
\en
where $c$ is an integration constant. In the case $c=0$, the universe is static and described by the Minkowski metric. In this case there is no singularity. For $c \neq 0$, we have
\be
\al = \pm \kappa^2 \varphi + {\bar c},
\label{s8}
\en
with ${\bar c}$ another integration constant. In terms of proper time $\tau$ for a co-moving observer (i.e.\ moving with the expansion of the universe), also called {\em cosmic proper time}, which is defined by $d\tau = N dt$, integration of \eqref{s7} yields $a=e^\al=\left[3(c\tau + {\tilde c}) \right]^{1/3}$, where ${\tilde c}$ is an integration constant, so that $a=0$ for $\tau = - {\tilde c}/c$ (and there is a big bang if $c>0$ and a big crunch if $c<0$). This means that the universe reaches the singularity in finite cosmic proper time.

In the usual quantum mechanical approach to the Wheeler-DeWitt theory, the complete description is given by the wave function and as such, as mentioned in the introduction, the notion of a singularity becomes ambiguous. Not so in the Bohmian theory. The Bohmian theory describes the evolution of an actual metric and hence there are singularities whenever this metric is singular, i.e., when $a=0$. The question of singularities in the special case where $V_M = V_G = 0$ was considered in \cite{pinto-neto12b,falciano15}. In this case, the Wheeler-DeWitt equation is
\be
\frac{1}{a^3}\pa^2_\varphi \psi - \kappa^2 \frac{1}{a^2} \pa_a (a \pa_a \psi) = 0,
\label{s9}
\en
or in terms of $\al$:
\be
\pa^2_\varphi \psi - \kappa^2 \pa^2_\al \psi = 0 .
\label{s10}
\en
The solutions are
\be
\psi = \psi_R(\ka \varphi - \al) + \psi_L(\ka \varphi + \al).
\label{s11}
\en
The actual metric might be singular; it depends on the wave function and on the initial conditions. For example, for a real wave function, $S=0$, the universe is static, so that there is no singularity. On the other hand, for wave functions $\psi=\psi_{R,L}$ the solutions are always classical, i.e., they are either static (if $\pa_\al S(\ka \varphi(0) - \al(0)) =0$, with $(\varphi(0),\al(0))$ the initial configuration) or they reach a singularity in finite cosmic proper time (if $\pa_\al S(\ka \varphi(0) - \al(0)) \neq 0)$. Wave functions with $\psi_R=-\psi_L$ satisfy $\psi(\varphi,\al) = -\psi(\varphi,-\al)$ and lead to trajectories that do not cross the line $\al =0$ in $(\varphi,\al)$-space. As such, trajectories starting with $\al(0) > 0$ will not have singularities. In this way bouncing solutions can be obtained. These describe a universe that contracts at early times then reaches a minimal volume and then expands again. At early and late times the evolution is classical.

Wave functions that have have no singularities are of the form
\be
\psi(\varphi,\al) = |\psi_R(\ka \varphi - \al)| + |\psi_L(\ka \varphi + \al)|\ee^{\ii \th}
\label{s12}
\en
(up to an irrelevant constant phase factor) with $\th$ a constant. The product $|\psi_R(\ka \varphi - \al)||\psi_L(\ka \varphi + \al)|$ is a constant of the motion in this case. For example, in the case $\psi_R(x)=\psi_L(x)=\ee^{-x^2}$, then $\al^2 + \varphi^2$ is constant and the solutions correspond to cyclic universes. In this case, we do not get classical behaviour at early or late times. 

In summary, there may or may not be singularities depending on the wave function and the initial conditions for the actual fields.

\subsubsection{The exponential potential}\label{exponential}
Consider $V_G = 0$ and the exponential matter potential
\begin{equation}
\label{def_pot}
V_M(\varphi) = V_0 \e^{-\lambda {\bar \kappa} \varphi},
\end{equation}
where $V_0$ and $\lambda$ are constant. ${\bar \kappa}^2 = 6 \kappa^2 = 8\pi G$, so that $\lambda$ is dimensionless. Such potentials have been widely explored in cosmology in order to describe in a simple way primordial inflation (which describes an exponential expansion of the universe driven by the inflaton field), the present acceleration of the universe, and matter bounces (which concern bouncing cosmologies with an initial matter-dominated phase of contraction). This is because they contain attractor solutions where the ratio between the pressure and the energy density is constant, $p/\rho = w$, with $w=(\lambda^2-3)/3$. In order to describe primordial or late accelerated expansion, one should have $-1/3 > w \geq -1$, and for matter bounces  $w\approx 0$, or $\lambda \approx \sqrt{3}$. Here we will discuss in detail the latter case.

The classical dynamics of such models is very rich and simple to understand. Assuming the gauge $N=1$ (so that the time is actually cosmic proper time) and defining the variables
\begin{equation}\label{mud1}
x = \frac{{\bar \kappa} }{\sqrt{6}H}\dot{\varphi}, \qquad
y = \frac{{\bar \kappa} \sqrt{V_M}}{\sqrt{3}H},
\end{equation}
where
\be
H=\frac{\dot a}{a} = {\dot \al}
\en
is the Hubble parameter, reduces the dynamical equations to
\begin{equation}\label{sist1c}
\frac{\dd x}{\dd \alpha} = -3 \left(x-\frac{\lambda}{\sqrt{6}}\right) \left(1-x\right)\left(1+x\right)
\end{equation}
and the Friedmann equation to
\begin{equation}
\label{xyFried}
x^2 + y^2 = 1.
\end{equation}
The ratio $w = p/\rho$ reads
\begin{equation}
w = 2x^2-1. \label{x_y_con}
\end{equation}

\begin{table}
\centering
\begin{tabular}{|c|c|c|}
\hline
$x$ & $y$ & $w$ \\
\hline
$-1$ & $0$ &  $1 $\\
\hline
$1$ & $0$  & $ 1$ \\
\hline
$\frac{\lambda }{\sqrt{6}}$ &$ -\sqrt{1-\frac{\lambda ^2}{6}}$ &  $\frac{1}{3} \left(\lambda ^2-3\right)$ \\
\hline
$\frac{\lambda }{\sqrt{6}}$ & $\sqrt{1-\frac{\lambda ^2}{6}}$  & $\frac{1}{3} \left(\lambda ^2-3\right)$ \\
\hline
\end{tabular}
\caption{Critical points of the planar system defined by \eqref{sist1c} and \eqref{xyFried}.}
\label{tab_crit}
\end{table}

As we are interested in investigating matter bounces, we will from now on set $\lambda = \sqrt{3}$.

The critical points are very easily identified from \eqref{sist1c}. They are listed in Tab.~\ref{tab_crit}. The critical points are $x=\pm 1$ with $w=1$ ($p=\rho$, stiff matter) and correspond to the space-time singularity $a=0$. Around this region, the potential is negligible with respect to the kinetic term. The critical points $x=1/\sqrt{2}$ with $w=0$ ($p=0$, dust matter)
are attractors (repellers) in the expanding (contracting) phase. Asymptotically in the infinite future (past) they correspond to very large slowly expanding (contracting) universes, and the space-time is asymptotically flat. Note that at $x=0$ the scalar field behaves like dark energy, $w=-1$, $p=-\rho$.

\begin{figure}[ht]
\centering
\includegraphics[width=0.5\textwidth]{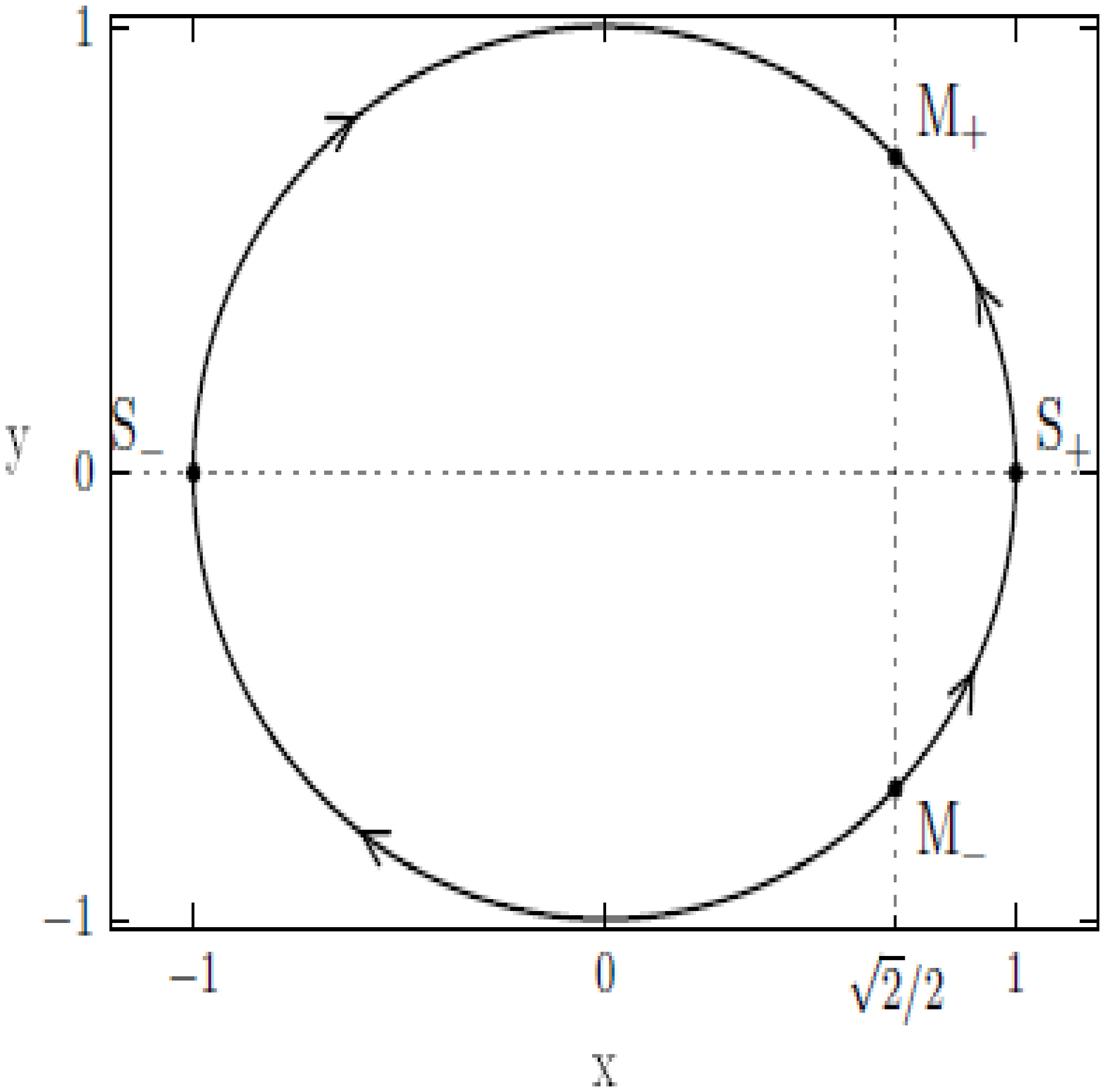}
\caption{Phase space for the planar system defined by \eqref{sist1c} and \eqref{xyFried}. The critical points are indicated by $M_\pm$ for a
matter-type effective equation of state, and $S_\pm$ for a
stiff-matter equation of state. For $y < 0$ we have the contracting phase, and for
$y > 0$ the expanding phase. Lower and upper quadrants are not physically
connected, because there is no classical mechanism that could drive a
bounce between the contracting and expanding phases: there is a
singularity in between.} \label{phase-classical}
\end{figure}

Hence we have four possible classical pictures:
\begin{itemize}
\item[a)]
A universe undergoing a classical dust contraction from very
large scales, the initial repeller of the model, and ending in a big crunch singularity around
stiff matter domination with $x\approx 1$, without ever passing through a dark energy phase.
\item[b)]
A universe undergoing a classical dust contraction from very
large scales, the initial repeller of the model, passing through a dark energy phase, and
ending in a big crunch singularity around
stiff matter domination with $x\approx -1$.
\item[c)]
A universe emerging from a big bang singularity around
stiff matter domination, with $x\approx 1$, and expanding to an asymptotically dust matter domination phase,
without ever passing through a dark energy phase (which is the time-reversed of case a).
\item[d)]
A universe emerging from a big bang singularity around
stiff matter domination, with $x\approx -1$, passing through a dark energy phase, and expanding to
an asymptotically dust matter domination phase (which is the time-reversed of case b).
\end{itemize}
These classical possibilities are depicted in Fig.~\ref{phase-classical}. The trajectories take place
on a circle. The points $M_{\pm}$ are respectively the dust attractor and repeller, while
$S_{\pm}$ are the singularities: the upper semi-circle is disconnected from the down semi-circle,
and they respectively describe the expanding and contracting solutions.

\begin{figure}[ht]
\centering
\includegraphics[width=0.5\textwidth]{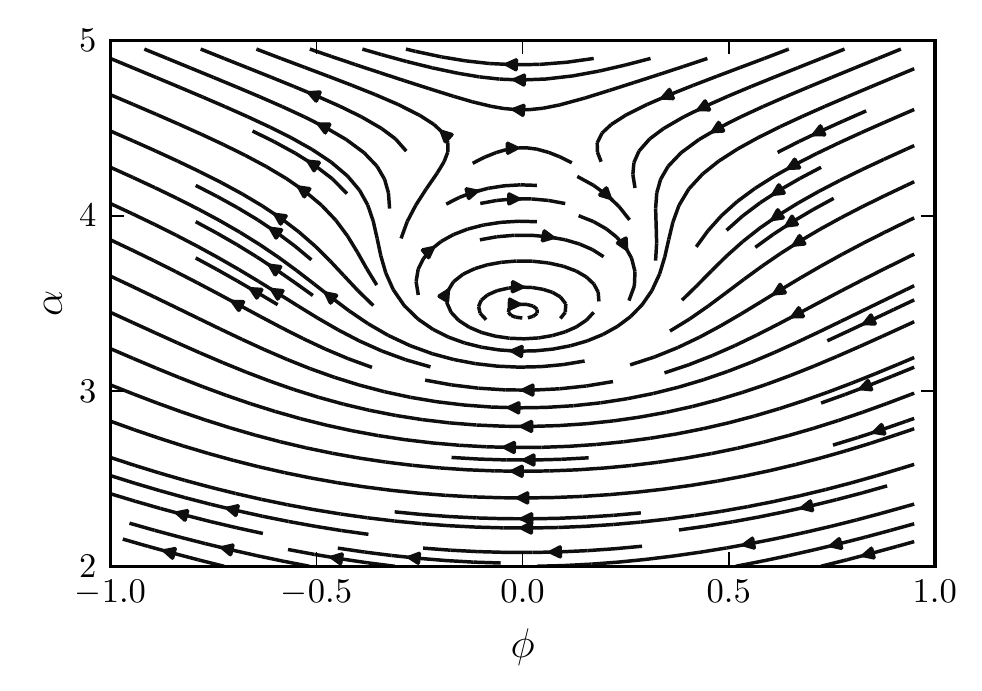}
\caption{Phase space for the quantum bounce \cite{bacalhau17}.  We can notice bounce and
cyclic solutions. The bounces in the figure correspond to case B, where
$\dot{\varphi}<0$, and it connects regions around $S_+$ in the contracting phase with
regions around $S_-$ in the expanding phase.} \label{quantum-bounce}
\end{figure}

In the quantum case, Bohmian bounce solutions were found. Exact solutions were given in \cite{colin17} and with some approximation in \cite{bacalhau17}, yielding the same qualitative picture. With these solutions, the quantum effects become important near the singularity. In this region, the potential is negligible and the quantum bounce is similar to the ones described in the preceding section or as in~\cite{colistete00}. The trajectories around the bounce are depicted in Fig.~\ref{quantum-bounce}. For large scale factors, $\alpha \gg 1$, the classical stiff matter behavior is recovered, $x \approx \pm 1$, and from there on the Bohmian trajectories become classical, as described above.

\begin{figure}[ht]
\centering
\includegraphics[width=0.5\textwidth]{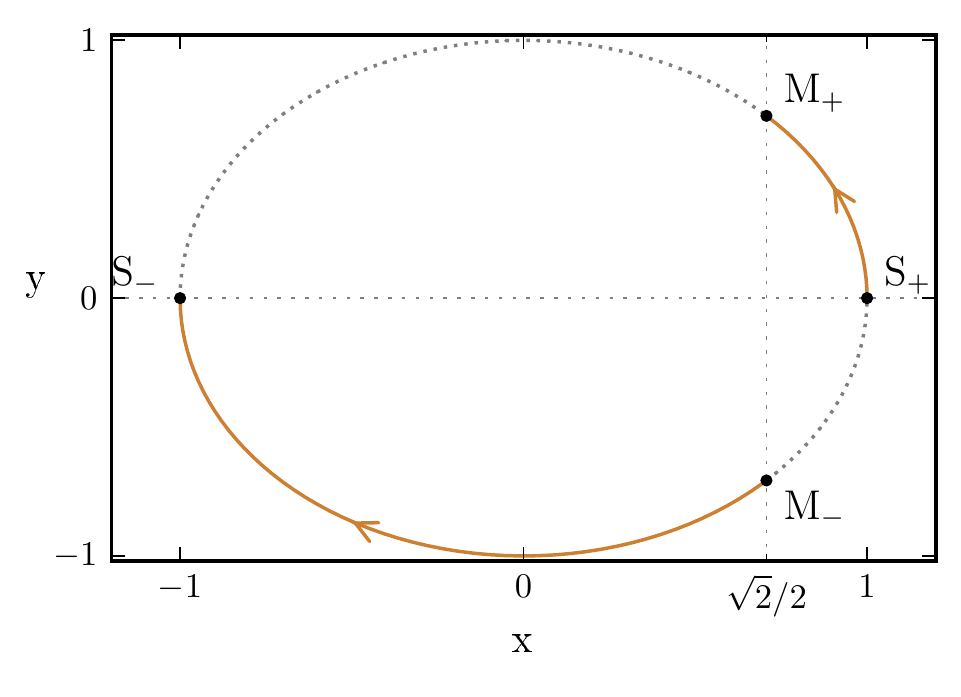}
\caption{Case A: the scalar field has a dark energy type equation of state during
contraction. By means of the quantum bounce, this system cannot address
the dark energy in the future, since the matter attractor is reached before.}
\label{phase-quantum-A}
\end{figure}

\begin{figure}[ht]
\centering
\includegraphics[width=0.5\textwidth]{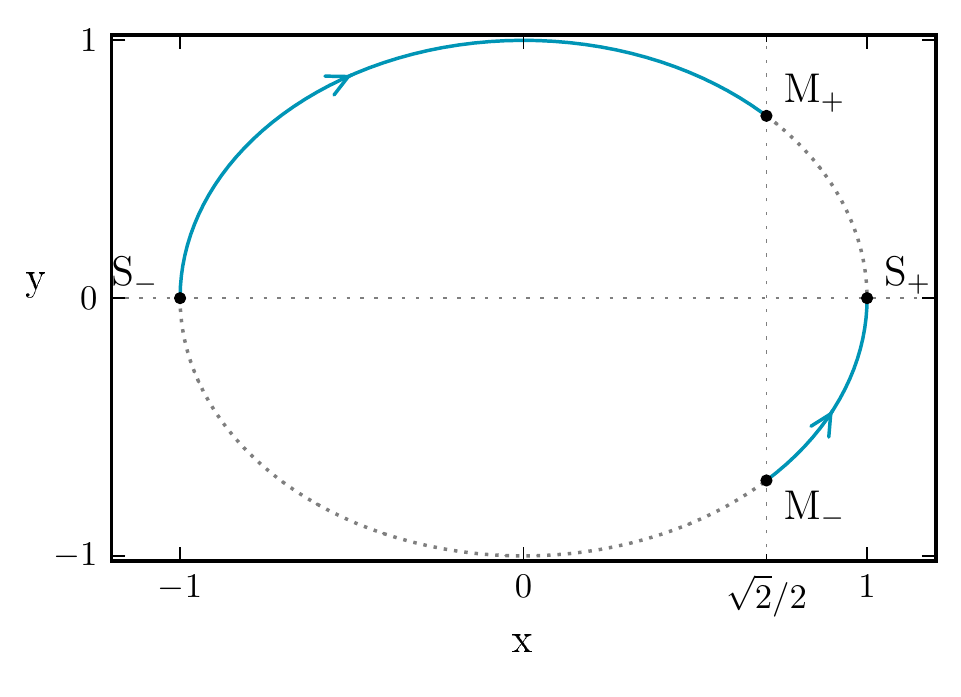}
\caption{Case B: the contracting phase begins close to the unstable point
$M_+$, in which the scalar field has a matter-type equation of state. After the quantum
bounce, the system emerges from $S_-$ and follows a dark energy phase until
reaches the future attractor $M_+$.} \label{phase-quantum-B}
\end{figure}

One very important observation is that, looking at Fig.~\ref{quantum-bounce}, the bounce can only connect
$x\approx\pm 1$ classical stiff matter domination regions with $x\approx\mp 1$ classical stiff matter
regions, respectively. In fact, a phase space analysis shows that such a connection of classical phases must happen for any bounce that might occur in the present model \cite{colin17,bacalhau17}. This fact implies that there are only two possible bouncing scenarios, see Figs.~\ref{phase-quantum-A} and \ref{phase-quantum-B}:
\begin{itemize}
\item[A)]
A universe undergoing a classical dust contraction from very
large scales, which passes through a dark energy phase before reaching a
stiff matter contracting phase with $x\approx -1$. In this regime, quantum effects become
relevant and a bounce takes place, launching the universe to a
classical stiff matter expanding phase with $x\approx 1$, which then evolves to an asymptotically
dust matter expanding phase, without passing through a dark energy phase.
\item[B)]
A universe undergoing a classical dust contraction from very
large scales, which goes to a
stiff matter contracting phase with $x\approx 1$,
without passing through a dark energy phase. In this regime, quantum effects become
relevant and a bounce takes place, launching the universe into a
classical stiff matter expanding phase with $x\approx -1$, which passes through a dark energy phase before reaching
an asymptotically dust matter expanding phase.
\end{itemize}

\begin{figure}[ht]
\centering
\includegraphics[width=0.5\textwidth]{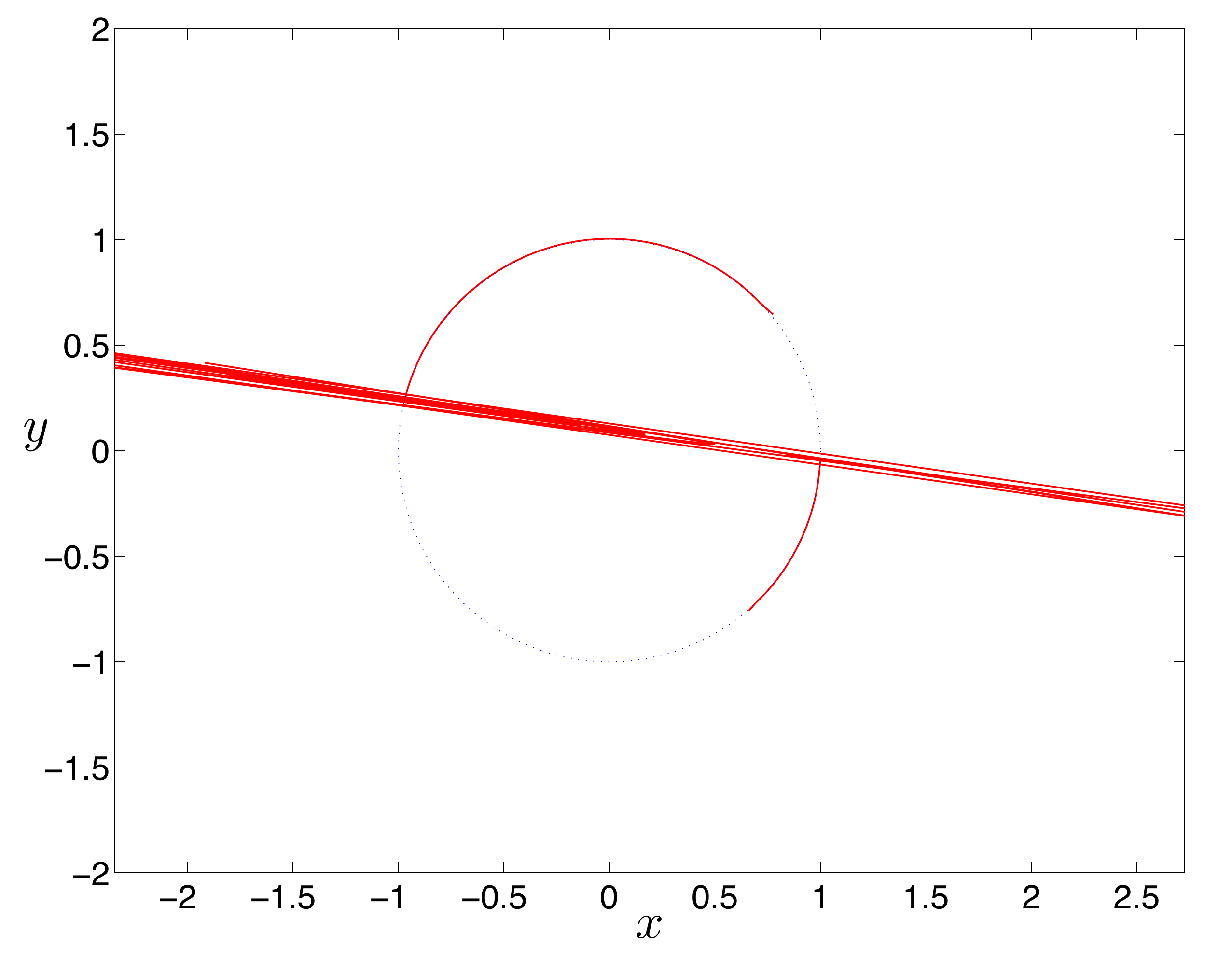}
\caption{Bohmian trajectory corresponding to an exact solution \cite{colin17}. It starts in the neighborhood of $(1/\sqrt{2}, -1/\sqrt{2})$ and ends in the neighborhood of $(1/\sqrt{2}, 1/\sqrt{2})$. The classical dynamics is valid almost everywhere, except near the singularity, where quantum effects become important and a bounce takes place, and the classical constraint $x^2 + y^2 = 1$ ceases to be satisfied.} \label{quantum-bounce2}
\end{figure}

Case B is the most physical one, because it can potentially describe the present observed acceleration of the universe as long as a dark energy era takes place in the expanding phase. Fig.\ \ref{quantum-bounce2} shows an example of an exact Bohmian trajectory. Note that it satisfies almost everywhere the classical constraint $x^2 + y^2 = 1$, except near the singularity, where the quantum bounce takes place, and the trajectory goes from the region $x\approx -1$ to the region where $x\approx 1$.

In section \ref{perturbations}, we return to this bouncing model and we analyze the evolution of perturbations on these backgrounds. This leads to a promising alternative to inflation.

\subsection{Mini-superspace - perfect fluid}
Another example of a mini-superspace model is that of a FLRW space-time with a perfect fluid, where the pressure is always proportional to the energy density, i.e., $p = w\rho$ with $w$ constant. This kind of fluid may describe the hot universe. Namely, at high temperatures, fields and particles become highly relativistic, with a radiation equation of state $p \approx \rho/3$. We will see again that Bohmian mechanics gives rise to non-singular solutions.

A perfect fluid can be modelled by a scalar field as follows. Consider the matter Lagrangian
\begin{equation}
\label{lg1}
L_M = \sqrt{-g} X^n ,
\end{equation}
where
\begin{equation}
\label{X}
X =  \frac{1}{2}g^{\mu\nu}\partial_{\mu}\varphi\partial_{\nu}\varphi .
\end{equation}
We will assume that $X \geq 0$ and we will interpret $\varphi$ as the potential yielding the normalized $4$-velocity of the fluid
\begin{equation}
\label{velocity}
V_{\mu} = \frac{\partial_{\mu}\varphi}{(2X)^{1/2}}.
\end{equation}
The energy-momentum tensor is given by
\be
T_{\mu\nu} = \frac{2}{\sqrt{-g}}\frac{\partial L_M}{\partial g^{\mu\nu}} =  2nX^nV_{\mu}V_{\nu}- g_{\mu\nu}X^n .
\en
Comparing with the usual expansion of the energy-momentum tensor in terms of energy density and pressure,
\begin{equation}
T_{\mu\nu} =   (\rho + p) V_{\mu}V_{\nu} - p g_{\mu\nu},
\end{equation}
we get
\begin{equation}
p = X^n , \qquad p = \frac{1}{2n-1}\rho,
\end{equation}
implying that $w = 1/(2n-1)$.

Assuming homogeneity, the scalar field depends only on time. The construction of the Hamiltonian is straightforward. The matter part reads
\be
H_M = c N \frac{p^{1+w}_\varphi}{a^{3w}}   ,
\en
where $p_{\varphi}$ is the momentum conjugate to $\varphi$ and $c= 1/w(\sqrt{2}n)^{1+w}$ is a constant. In the case of $w=1$, the matter Hamiltonian is that of the previous section. Before applying canonical quantization, the following canonical transformation is performed:
\begin{equation}
T= \frac{1}{c(1+w)} \frac{\varphi}{ p_\varphi^{w}} , \qquad P_{T}= c p_\varphi^{1+w},
\label{can}
\end{equation}
so that
\begin{equation}
H_{M} = N\frac{P_{T}}{a^{3w}}.
\label{h1}
\end{equation}
An important property is that the momentum now appears linearly. Combining this perfect fluid Hamiltonian with the gravitational Hamiltonian for a FLRW geometry, the total mini-superspace Hamiltonian is obtained:{\footnote{In this section, we follow the notation of \cite{pinto-neto13}, where units are such that $\kappa^2 = 1/2$. (Compared to the previous section the total Lagrangian was also divided by $\ka^2$.)}}
\begin{equation}
H=N\left(-\frac{P_a^{2}}{4a} + \frac{P_{T}}{a^{3w}}\right).
\label{h2}
\end{equation}
It implies that ${\dot T} = N/a^{3w}$ or in terms of cosmic proper time $\tau$, $dT/d\tau =1/ a^{3w}$ and hence $T$ increases monotonically, so that it can be used as a clock variable. In terms of $T$, the scale factor evolves like $a\propto T^{2/3(1-w)}$ in the case $w \neq 1$, which is singular at $T=0$ (if the proportionality constant is different from zero).

In the quantum case, because one momentum appears linearly in the Hamiltonian, the Wheeler-DeWitt equation assumes the Schr\"odinger form \cite{alvarenga98,acacio98,alvarenga02}
\begin{equation}
\ii \frac{\partial}{\partial T}\Psi(a,T) =\frac{1}{4} \left\{
a^{(3\omega-1)/2}\frac{\partial}{\partial a} \left[
a^{(3\omega-1)/2}\frac{\partial}{\partial a}\right]
\right\}\Psi(a,T).
\end{equation}
Note that in the case $w=1$, this equation differs from the Wheelder-DeWitt equation \eqref{s9}, due to the different pair of canonical variables which were quantized. In the rest of this section, we will only consider $w \neq 1$ (for these cases we can apply the transformation \eqref{transf}). The guidance equations are
\be
\label{guidancec} {\dot T} = \frac{N}{a^{3w}} , \qquad \dot{a}=-\frac{N}{2a} \frac{\partial S}{\partial a}.
\en
The dynamics can be simplified using the transformation
\be
\label{transf}
\chi=\frac{2}{3} (1-\omega)^{-1} a^{3(1-\omega)/2},
\en
to obtain
\begin{equation}
\ii\frac{\partial\Psi(a,T)}{\partial T}= \frac{1}{4} \frac{\partial^2\Psi(a,T)}{\partial \chi^2} \label{es202}.
\end{equation}
This is just the time-inversed Schr\"odinger equation for a one-dimensional free particle with mass $2$ constrained to the positive axis.

In the context of orthodox quantum theory, the form of the Wheeler-DeWitt equation suggest to interpret $T$ as time and to find a corresponding suitable Hilbert space. Since $\chi>0$, the Hilbert space requires a boundary condition
\begin{equation}
\label{cond27}
\Psi \bigl|_{\chi=0} = c \frac{\partial\Psi}{\partial
\chi}\Biggl|_{\chi=0},
\end{equation}
with $c \in \mathbb{R} \cup \{\infty\}$. $|\Psi^2| d\chi$ is then the probability measure for the scale factor. The boundary condition ensures that the total probability is preserved in time.

Note, however, that even though this form is suggestive, it is still rather ad hoc to interpret $T$ as time. For example other variables could have been chosen (in particular if extra matter fields were considered). As explained before, in the Bohmian theory, the time $t$ is unobservable and the physical clocks should be modeled by field or metric degrees of freedom. Since $T$ increases monotonically with $t$, as long as the singularity $a=0$ is not obtained, it can be used as a clock variable. But other monotonically increasing variables could also be used as clocks without ambiguities. The dynamics of the scale factor can be expressed in terms of $T$:
\be
\frac{da}{dT} = - \frac{a^{3w-1}}{2} \frac{\pa S}{\pa a}
\en
or
\be
\frac{d \chi}{dT} = - \frac{1}{2} \frac{\pa S}{\pa \chi}.
\en
In the Bohmian approach, the condition \eqref{cond27} implies that there are no singularities \cite{pinto-neto05b} (because the condition means that there is no probability flux $J_\chi \sim {\textrm{Im}}\left(\Psi^* \frac{\pa \Psi}{\pa \chi}\right)$ through $\chi=0$, so no trajectories will cross $a=0$). However, for wave functions not satisfying the boundary condition \eqref{cond27}, singularities will be obtained at least for some trajectories. For example, for a plane wave, the trajectories are the classical ones and hence a singularity is always obtained. From the Bohmian point of view this can motivate the consideration of a Hilbert space based on \eqref{cond27}. It is then also natural to use $|\Psi^2| d\chi$ as the normalizable equilibrium distribution for the scale factor.

As an example of a wave function that satisfies the condition (\ref{cond27}), consider the Gaussian
\begin{equation}
\label{initial}
\Psi^{(\mathrm{init})}(\chi)=\biggl(\frac{8}{T_0\pi}\biggr)^{1/4}
\exp\left(-\frac{\chi^2}{T_0}\right) ,
\end{equation}
where $T_0$ is an arbitrary constant. The wave solution for all times in terms of $a$ is \cite{acacio98,alvarenga02}:
\begin{multline}\label{psi1t}
\Psi(a,T)=\left[\frac{8 T_0}{\pi\left(T^2+T_0^2\right)}
\right]^{1/4}
\exp\biggl[\frac{-4T_0a^{3(1-\omega)}}{9(T^2+T_0^2)(1-\omega)^2}\biggr]
\nonumber\\
\times\exp\left\{-\ii\left[\frac{4Ta^{3(1-\omega)}}{9(T^2+T_0^2)(1-\omega)^2}
+\frac{1}{2}\arctan\biggl(\frac{T_0}{T}\biggr)-\frac{\pi}{4}\right]\right\}.
\end{multline}
The corresponding Bohmian trajectories are
\begin{equation}
\label{at} a(T) = a_0
\left[1+\left(\frac{T}{T_0}\right)^2\right]^\frac{1}{3(1-\omega)} .
\end{equation}
Note that this solution has no singularities for any initial value of $a_0 \neq 0$, and tends to the classical solution when $T\rightarrow\pm\infty$. The solution (\ref{at}) can also be obtained for other initial wave functions \cite{alvarenga02}.

For $w=1/3$ (radiation fluid), and adjusting the free parameters, the solution \eqref{at} can reach the classical evolution before the nucleosynthesis era, where the standard cosmological model starts to be compared with observations. Hence, it can be a good candidate to describe a sensible cosmological model at the radiation dominated era which is free of singularities.

\subsection{Loop quantum cosmology}
Loop quantization is a different way to quantize general relativity \cite{rovelli04,rovelli14}.  Application of  this quantization method to the classical mini-superspace model defined by \eqref{s2} results in the following theory. States are functions $\psi_\nu(\varphi)$ of a continuous variable $\varphi$ and a discrete variable
\be
\nu = \epsilon C a^3,
\label{s15}
\en
with
\be
C = \frac{ V_0}{2\pi G \gamma },
\label{s16}
\en
where $\epsilon = \pm 1$ is the orientation of the triad (which is used in passing from the metric representation of general relativity to the connection representation), $V_0$ is the fiducial volume (which is introduced to make volume integrations finite) and $\gamma$ is the Barbero-Immirzi parameter. $\nu$ is discrete as it is given by $\nu = 4 n\lambda$ with $n \in {\mathbb Z}$ and $\lambda^2 = 2 \sqrt{3}\pi\gamma G$. The value $\nu=0$, which corresponds to the singularity, is included. One could also take $\nu = \epsilon + 4 n\lambda$, with $\epsilon \in (0,4\lam)$. This does not include the value $\nu=0$ and as such the singularity would automatically be avoided in the corresponding Bohmian theory (because, as will be discussed, in the Bohmian theory the possible values the scale factor can take are given by the discrete values of $\nu$ on which $\psi$ has its support).

As usual, the quantization is not unique. Because of operator ordering ambiguities, different wave equations may be obtained. Different operator orderings are considered in the literature \cite{ashtekar06d,ashtekar08,martin-benito09,mena-marugan11,banerjee12}. In all models, the wave equation is of the form
\be
B_\nu \pa^2_\varphi \psi_\nu(\varphi) + \sum_{\nu'} K_{\nu,\nu'} \psi_{\nu'}(\varphi) = 0 ,
\label{s17}
\en
with $\psi_\nu =\psi_{-\nu} $ and $B_\nu$ and $K_{\nu,\nu'}=K_{\nu',\nu}$ are real. The gravitational part, determined by $K$, is not a differential equation but a difference equation. For example, in the APS model \cite{ashtekar06d,ashtekar08}, the wave equation is
\be
B_\nu \pa^2_\varphi \psi_\nu(\varphi)  - 9 \kappa^2 D_{2\lam} (|\nu| D_{2\lam} \psi_\nu(\varphi)) = 0,
\label{s18}
\en
where
\be
D_h \psi_\nu = \frac{\psi_{\nu + h} - \psi_{\nu - h}}{2h},
\label{s19.0}
\en
so that
\be
K_{\nu,\nu\pm 4\lam} = - \frac{9 \kappa^2}{16\lam^2} |\nu \pm 2 \lam| \,, \qquad   \qquad  K_{\nu,\nu} = - K_{\nu,\nu+4\lam} -   K_{\nu,\nu-4\lam}
\label{s19}
\en
and the other $K_{\nu,\nu'}$ are zero. Various choices for $B_\nu$ exist, again due to operator ordering ambiguities \cite{bojowald02,bojowald08}. One choice is \cite{ashtekar08}:
\be
B_\nu =  \left|\frac{3}{2}D_\lam|\nu |^{2/3} \right|^3 = \left|\frac{3}{2} \frac{|\nu + \lam|^{2/3} - |\nu - \lam|^{2/3} }{2\lam}\right|^3.
\label{s20}
\en
All choices of $B_\nu$ in all the models (except in the simplified APS model \cite{ashtekar08}, called sLQC) share the important properties that $B_0=0$ and that for $|\nu| \gg \lam$ (taking the limit $\lam \to 0$, or equivalently, taking the Barbero-Immirzi parameter or the area gap to zero), $B_\nu \to 1/|\nu|$. For $|\nu| \gg \lam$ (taking the limit $\lam \to 0$), this wave equation reduces to the Wheeler-DeWitt equation
\be
\frac{1}{|\nu|}\pa^2_\varphi \psi - 9\kappa^2 \pa_\nu (|\nu| \pa_\nu \psi) = 0 ,
\label{s21}
\en
which is just the wave equation \eqref{s10} in terms of $\nu$.

Since the gravitational part of the wave equation \eqref{s17} is now a difference operator, rather than a differential operator, the Bohmian dynamics now concerns a jump process rather than a deterministic process. Such processes have been introduced in the context of quantum field theory to account for particle creation and annihilation \cite{bell84,duerr032,duerr05a}. In the Bohmian theory, the scalar field evolves continuously, while the scale factor $a$, which will be expressed in terms of $\nu$ using \eqref{s15}, takes discrete values, determined by $\nu = 4 n\lambda$ with $n \in {\mathbb Z}$. Since the evolution of the scale factor is no longer deterministic, but stochastic, the metric is no longer Lorentzian. Namely, once there is a jump, the metric becomes discontinuous. The metric is only ``piece-wise''  Lorentzian, i.e., Lorentzian in between two jumps.

The Bohmian dynamics can be found by considering the continuity equation, which follows from \eqref{s17}:
\be
\pa_\varphi J_{\nu}(\varphi)  = \sum_{\nu'} J_{\nu,\nu'}(\varphi) ,
\en
where
\be
J_{\nu}(\varphi) = B_\nu \pa_\varphi S_\nu(\varphi) ,\qquad   J_{\nu, \nu'}(\varphi) = -K_{\nu,\nu'} \textrm{Im}\left( \psi_\nu (\varphi)\psi^*_{\nu'} (\varphi)\right).
\en
$J_{\nu, \nu'}$ is anti-symmetric and non-zero only for $\nu'=\nu \pm 4\lam$ for the LQC models considered above. Writing
\be
\sum_{\nu'} J_{\nu,\nu'} = \sum_{\nu'} \left( {\widetilde T}_{\nu, \nu'} |\psi_{\nu'}|^2 - {\widetilde T}_{\nu',\nu} |\psi_{\nu}|^2\right),
\en
where
\begin{equation}
{\widetilde T}_{\nu, \nu'}(\varphi) = \left\{
\begin{array}{ll}
\frac{J_{\nu, \nu'}(\varphi)}{|\psi_{\nu'}(\varphi)|^2} & \quad \text{if } J_{\nu, \nu'} (\varphi)> 0\\
0 &\quad  \text{otherwise}
\end{array} \right.,
\end{equation}
we can introduce the following Bohmian dynamics which preserves the quantum equilibrium distribution $|\psi_{\nu}(\varphi)|^2d\varphi$. The scalar field satisfies the guidance equation
\be
\dot \varphi = NC B_\nu \pa_\varphi S_\nu ,
\label{s22}
\en
where $\psi_\nu = |\psi_\nu|\ee^{\ii S_\nu }$. The variable $\nu$, which determines the scale factor, may jump $\nu' \to \nu$ with transition rates given by $T_{\nu, \nu'}(\varphi) = NC {\widetilde T}_{\nu, \nu'}(\varphi)$. That is, $T_{\nu, \nu'}(\varphi)$ is the probability to have a jump $\nu' \to \nu$ in the time interval $(t,t+dt)$. Note that the jump rates at a certain time depend on both the wave function and on the value of $\varphi$ at that time. The properties of $J_{\nu, \nu'}$ imply that for a fixed $\nu$ either $T_{\nu, \nu +4\lam}$ or $T_{\nu, \nu -4\lam}$ may be non-zero (not both). The jump rates are ``minimal'', i.e., they correspond to the least frequent jump rates that preserve the quantum equilibrium distribution \cite{duerr05a}. Just as in the classical case and the Bohmian Wheeler-DeWitt theory, the lapse function is arbitrary, which guarantees time-reparameterization invariance, just as in the case of Wheeler-DeWitt quantization. For $|\nu| \gg \lam$ (taking the limit $\lam \to 0$), this Bohmian theory reduces to the one of the Wheeler-DeWitt equation (using similar arguments as in \cite{vink93}).

Let us now turn to the question of singularities. If $T_{0, \pm 4\lambda} = 0$, then the scale factor $a$ (or $\nu$) can never jump to zero, so a big crunch is not possible. If $T_{\pm 4\lambda, 0} = 0$, then the scale factor can not jump from zero to a non-zero value, so a big bang is not possible. Hence there are no singularities if $J_{0, \pm 4\lambda} = 0$. That this condition is satisfied can be seen as follows. Since $B_0=0$, we have
\be
K_{0,4\lam} \psi_{4\lam} + K_{0,-4\lam} \psi_{-4\lam} +  K_{0,0} \psi_{0} = 0.
\en
Using the properties $K_{0,\nu} = K_{0,-\nu}$ and $\psi_\nu = \psi_{-\nu}$, we obtain that
\be
\textrm{Im}\left( \psi^*_0 K_{0,\pm 4\lam} \psi_{\pm 4\lam} \right) = 0
\en
and hence that $J_{0, \pm 4\lambda} = 0$. In summary, Bohmian loop quantum cosmology models for which the wave equation \eqref{s17} has the properties that $B_0=0$, $K_{0,\nu} = K_{0,-\nu}$ and $\psi_\nu = \psi_{-\nu}$, do not have singularities. Importantly, no boundary conditions need to be assumed.

In the case that $\psi$ is real, both $\varphi$ and $a$ are static. For other possible solutions, the wave equation needs to be solved first. This is rather hard, but can perhaps be done in the simplified model sLQC since the eigenstates of the gravitational part of the Hamiltonian are known in this case. Something can be said about the asymptotic behaviour however. Since for large $\nu$ this Bohmian theory reduces to the Bohmian Wheeler-DeWitt theory, the trajectories will tend to be classical in this regime. Namely consider solutions \eqref{s11} to the Wheeler-DeWitt equation for which the functions $\psi_R$ and $\psi_L$ go to zero at infinity. Then for $\al \to \infty$, the wave functions $\psi_R$ and $\psi_L$ become approximately non-overlapping in $(\varphi,\al)$-space. As such the Bohmian motion will approximately be determined by either $\psi_R$ or $\psi_L$ and hence classical motion is obtained. This implies an expanding or contracting (or static) universe. We expect that a bouncing universe will be the generic solution.

So far we assumed $k=\Lambda=0$. In the case $k=\pm 1$ or $\Lambda \neq 0$ singularities are also eliminated \cite{struyve17b}.

In conclusion, in Bohmian loop quantum cosmology, there is no big bang or big crunch singularity regardless of the wave function. The result follows from a very simple dynamical analysis. It is in agreement with the results derived in the standard quantum mechanical framework \cite{ashtekar06b,ashtekar06c,ashtekar06d}. However, in \cite{ashtekar06b,ashtekar06c,ashtekar06d}, $\varphi$ is considered a time variable from the start, whereas in the Bohmian case, $\varphi$ can only be used as a clock variable when it increases monotonically with $t$. In addition, often only a special class of wave functions is considered, namely the ones that behave classically at ``late times''.

\section{Cosmological perturbations}\label{perturbations}
In section \ref{singularities}, we have described Bohmian mini-superspace models. These simplified models of quantum gravity were obtained by assuming homogeneity and isotropy. In this section, we consider deviations from homogeneity and isotropy by introducing perturbations. These perturbations are very important in current cosmological models, either inflationary or bouncing models, because they form the seeds of structure formation. Namely, according these models, in the far past the universe was so homogeneous that the only sources of inhomogeneities were quantum vacuum fluctuations. During the  subsequent expansion of the universe the vacuum fluctuations result in classical fluctuations of the matter density. The classical fluctuations then grow through gravitational clumping and give rise to structures such as galaxies and clusters of galaxies we observe today. These vacuum fluctuations also leave an imprint on the cosmic microwave background radiation as temperature fluctuations.

There are a number of issues with this standard account that the Bohmian approach helps to solve. First of all, the conventional approach to deal with the cosmological perturbations is to consider a semi-classical treatment where only the first-order perturbations are quantized, while the background is treated classically (without back-reaction from perturbations onto fluctuations). This was largely explored in inflationary models in order to calculate the primordial power spectrum of scalar and tensor cosmological perturbations, and evaluate their observational consequences. However, the classical treatment of the background implies that there is a singularity, a point where no physics is possible, rendering the analysis incomplete. Using Bohmian mechanics, the usual approach to cosmological perturbations can be extended to include quantum corrections to the background evolution. This can then be used to infer consequences for the formation of structures in the universe, and for the anisotropies of the cosmic background radiation. Early attempts on this approach resulted in very complicated and intractable equations \cite{halliwell86}. Using Bohmian mechanics, one is able to tremendously simplify the evolution equations of quantum cosmological perturbations in quantum backgrounds, rendering them into a simple and solvable form, suitable for the calculation of their observational consequences \cite{peter05,peter06,peter07,pinho07,vitenti12,vitenti13,falciano13,pinto-neto14b}. We start with illustrating the derivation of the motion of the quantum perturbations in a quantum background in section \ref{quantumbackground} for the simple case of a canonical scalar with zero potential. Similar results can be obtained for non-zero potentials. Then, in section \ref{observational}, we will discuss the observational consequences in the case of an exponential matter potential, for which the background equations yield bouncing solutions as discussed in section \ref{exponential}.

A second problem with the conventional approach is that of the quantum-to-classical transition of the perturbations \cite{perez06,sudarsky11}. Namely, the quantum vacuum fluctuations somehow turn into classical fluctuations during the evolution of the universe. But this is difficult to account for in the context of orthodox quantum theory. We will consider this in a bit more detail for the case of inflation theory in section \ref{qtoc}, for bounce theories see \cite{pinto-neto14a}.

\subsection{Cosmological perturbations in a quantum cosmological background}\label{quantumbackground}

The mini-superspace bouncing non-singular models described in section \ref{singularities} considered a hydrodynamical fluid or a scalar field as their matter contents. Here, we will present the main features for the quantum treatment of perturbations and background in the case of a canonical scalar field. We will consider a free scalar field, i.e., with zero field potential. The generalization to other potentials (like inflationary ones \cite{falciano13}) is straightforward. Hydrodynamical fluids are treated in \cite{peter05,peter06,peter07,pinho07}.

The free massless scalar field is $\varphi\left(t,x\right)=\varphi_0 \left(t\right) +\delta \varphi\left(t,x\right)$, where $\varphi_0$ is the background homogeneous scalar field and $\delta \varphi\left(t,x\right)$ is its linear perturbation. The FLRW metric together with its scalar perturbations is given by
\begin{equation}
\label{perturb}
g_{\mu\nu}=g^{(0)}_{\mu\nu}+h_{\mu\nu},
\end{equation}
where $g^{(0)}_{\mu\nu}$ represents a homogeneous and isotropic FLRW cosmological background,
\begin{equation}
\label{linha-fried}
 {\rm {d}} s^{2}=g^{(0)}_{\mu\nu} {\rm {d}} x^{\mu} {\rm {d}} x^{\nu}=N^{2}(t) {\rm {d}} t^2 -
a^{2}(t)\delta_{ij} {\rm {d}} x^{i} {\rm {d}} x^{j},
\end{equation}
where we assumed a flat spatial metric, and $h_{\mu\nu}$ represents linear scalar perturbations around it, which we decompose into
\begin{eqnarray}
\label{perturb-componentes}
h_{00}&=&2N^{2}\phi, \nonumber \\
h_{0i}&=&-NaB_{,i}, \\
h_{ij}&=&2a^{2}(\psi\gamma_{ij}-E_{,ij}), \nonumber
\end{eqnarray}
where we used the notation $B_{,i}=\pa_i B$. The case of tensor perturbations, i.e., gravitational waves, is very similar and actually easier \cite{peter05,peter06}.

Starting from the classical action for this system, the Hamiltonian up to second-order can be brought into the following simple form (using a redefinition of $N$ with terms which do not alter the equations of motion up to first order and performing canonical transformations), without ever using the background equations of motion \cite{falciano09} ($\ka^2=1$):
\begin{equation}
\label{h0}
H=\frac{N}{2 e^{3\alpha}}\left[-P_{\alpha}^{2} +P_{\varphi}^2+\int d ^3x \left( \frac{\pi^2}{\sqrt{\gamma}}+\sqrt{\gamma}e^{4\alpha}v^{,i}v_{,i}\right)\right] ,
\end{equation}
where we dropped the subscript 0 from the background field and where again $a=\ee^{\al}$ and $v({\bf x})$ is the usual Mukhanov-Sasaki variable \cite{mukhanov92}, defined as
\begin{equation}
\label{vinculo-simples2}
v = a\biggl(\delta\varphi + \frac{\varphi ' \phi}{\cal{H}}\biggr) ,
\end{equation}
with primes denoting derivatives with respect to conformal time $\eta$, defined by $d\eta = d\tau /a$, $\tau$ being cosmic proper time, and $\mathcal{H} = a'/a = \al'$.

This system is straightforwardly quantized and yields the Wheeler-DeWitt equation
\begin{equation}
\label{split-h0}
(\hat{H}_0+\hat{H}_2) \Psi = 0,
\end{equation}
where
\begin{eqnarray}
&&\hat{H}_0=-\frac{\hat{P}_\alpha^2}{2}+\frac{\hat{P}_\varphi^2}{2} , \\
&&\hat{H}_2=\frac{1}{2}\int d ^3x \left( \frac{\hat{\pi}^2}{\sqrt{\gamma}}+\sqrt{\gamma}e^{4\hat{\alpha}}\hat{v}^{,i}\hat{v}_{,i}\right) .
\end{eqnarray}

We now want to consider an approximation where the background evolves independently from the perturbations. The evolution of the background will be Bohmian rather than classical (as is usually considered). This approximation is obtained as follows. We write the wave function as
\be
\Psi(\al,\varphi,v) = \Psi_0(\al,\varphi)\Psi_2(\al,\varphi,v)
\en
and assume that $|\Psi_2| \ll |\Psi_0|$ and $|S_2| \ll |S_0|$, together with their derivatives with respect to the background variables. Then to lowest order we have
\be
\hat{H}_0 \Psi_0 =  0 ,
\en
and the usual corresponding guidance equations. This is the mini-superspace model described in section \ref{mcsf}. As we have seen, quantum effects can eliminate the background singularity leading to bouncing models.

Using a Bohmian solution $(\al(\eta),\varphi(\eta))$ for the background, guided by $\Psi_0$, an approximate wave equation for the perturbations can now be obtained. It is found by considering the conditional wave function
\be
\chi(v,\eta) = \Psi_2(\al(\eta),\varphi(\eta),v)
\en
for the pertubations. It approximately satisfies (after suitable transformations)
\begin{equation}
\label{xo}
\ii \frac{\partial \chi(v,\eta)}{\partial \eta}=\frac{1}{2}\int d^3x \left( \hat{\pi}^2 + \hat{v}^{,i}\hat{v}_{,i}-\frac{a''}{a} \hat{v}^2\right) \chi(v,\eta).
\end{equation}
This is the same wave equation for the perturbations known in the literature, in the absence of a scalar field potential \cite{mukhanov92}. When a scalar field potential is present, one just has to substitute $a''/a$ by $z''/z$ in this Hamiltonian, where $z = a\varphi '/{\cal{H}}$. The crucial difference with the standard account is that now the time-dependent potential $a''/a$ or $z''/z$ in Eq.~\eqref{mode-f2} can be rather different from the semi-classical one because it is calculated from Bohmian trajectories, not from the classical ones. This can give rise to different effects in the region where the quantum effects on the background are important, which can propagate to the classical region, yielding different observations.

\subsection{Bunch-Davies vacuum and power spectrum}\label{bdvacuum}
Having found the evolution equation for quantum perturbations in a quantum background, we recall the solution of interest in both inflationary and bouncing models, which is the Bunch-Davies vacuum.

Let us first apply the unitary transformation $\ee^{\ii \frac{z''}{z} \hat{v}^2}$ to \eqref{xo} (with $a''/a$ replaced by $z''/z$ to describe general potentials), which brings the Schr\"odinger equation into the form{\footnote{Both forms \eqref{xo} and \eqref{xo2} are commonly used in the literature.}}
\begin{equation}
\label{xo2}
\ii \frac{\partial \Psi(v,\eta)}{\partial \eta} = \frac{1}{2}\int d^3x \left[ \hat{\pi}^2 + \hat{v}^{,i}\hat{v}_{,i}+ \frac{z'}{z} \left( \hat{\pi}\hat{v}+ \hat{v}\hat{\pi}\right)\right] \Psi(v,\eta).
\end{equation}
Introducing the Fourier modes $v_{\bf k}$ of the Mukhanov-Sasaki variable, defined by
\begin{equation}
\label{mode}
v({\bf x})=\int{\frac{d^3x}{(2\pi)^{3/2}}v_{\bf k} \ee^{\ii {\bf k} \cdot {\bf x}}},
\end{equation}
and assuming a product wave function
\be
\label{product}
\Psi= \Pi_{{\bf k} \in \setR^{3+}} \Psi_{\bf k}(v_{\bf k},v^*_{\bf k},\eta),
\en
each factor $\Psi_{\bf k}$ satisfies the Schr\"odinger equation
\begin{equation}
\label{sch}
\ii\frac{\partial\Psi_{\bf k}}{\partial\eta}=
\left[ -\frac{\partial^2}{\partial v_{\bf k}^*\partial v_{\bf k}}+
k^2 v_{\bf k}^* v_{\bf k}
- \ii\frac{z'}{z}\left(\frac{\partial}{\partial v_{\bf k}^*}v_{\bf k}^*+
v_{\bf k}\frac{\partial}{\partial v_{\bf k}}\right)\right]\Psi_{\bf k}.
\end{equation}
The corresponding guidance equations are
\begin{equation}
\label{guidance}
v'_{\bf k}= \frac{\partial S_{\bf k}}{\partial v^*_{\bf k}}+\frac{z'}{z}v_{\bf k} .
\end{equation}

The Bunch-Davies vacuum is of the form \eqref{product}, with
\begin{equation}
\label{psi2}
\Psi_{\bf k} = \frac{1}
{\sqrt{\pi}|f_k(\eta)|} \exp{\left\{-\frac{1}{2|f_k(\eta)|^2}|v_{\bf k}|^2 +   \ii \left[\left(\frac{|f_k(\eta)|'}{|f_k(\eta)|}-
\frac{z'}{z}\right)|v_{\bf k}|^2-
\int^\eta \frac{d {\tilde \eta}}{2|f_k({\tilde \eta})|^2}\right]\right\}} ,
\end{equation}
with $f_k$ a solution to the classical mode equation
\begin{equation}
\label{mode-f}
f_{\bf k}'' +  \left(k^2-\frac{z''}{z}\right) f_{\bf k}=0,
\end{equation}
with initial conditions $f_k(\eta_i) = \exp{(-\ii k\eta_i)}/\sqrt{2k}$, at an early time $|\eta_i|\gg1$  when the physical modes satisfy $k^2 \gg z''/z$. This state is homogeneous and isotropic. The guidance equations are easily integrated and yield
\be
\label{soly}
v_{\bf k}(\eta) =  v_{\bf k}(\eta_i)\frac{|f_k(\eta)|}{|f_k(\eta_i)|}.
\en
This result is independent of the precise form of $f_k(\eta)$ and hence is quite general.

The Bunch-Davies vacuum is motivated as follows. In inflationary models and bouncing models, $z'/z \propto 1/|\eta|\approx 0$ at early times, i.e., for $|\eta|\gg 1$. Hence, in this limit, the quantum perturbations behave like a bunch of quantum mechanical harmonic oscillators and the Bunch-Davies vacuum tends to the vacuum state of the quantum harmonic oscillator. In the case of inflation theory, the inflaton field drives the universe to a homogeneous state so that only vacuum fluctuations of these pertubations remain. Similarly, in the case of a bouncing model, in the far past in, the matter content of the universe was homogeneously and isotropically diluted in an immensely large space which was slowly contracting. In this very mild matter contraction, space-time was almost flat and empty, and the only source of inhomogeneities could only be small quantum vacuum fluctuations.

In the next section, we discuss how this formalism connects to current cosmological observations. In section \ref{qtoc}, we discuss the quantum-to-classical transition of the perturbations and then finally, in section \ref{observational}, we discuss the cosmological perturbations for the matter bounce quantum background described in section \ref{exponential}. This approach models the realistic situation where an accelerated era takes place in the expanding phase. In addition to the scalar perturbations, we will also discuss the results for the case of primordial gravitational waves. As we shall see, the quantum bounce solves important problems which cannot be addressed by classical bounces, and yield observational imprints on the cosmic microwave background radiation.

\subsection{Power spectrum and cosmic microwave background}
To make the connection between the early universe and present cosmological observations, in particular the anisotropies of the cosmic microwave background, the quantity of interest is the two-point correlation function
\be
\label{twopoint}
\left\langle \hat{v}({\bf x},\eta) \hat{v}({\bf x} + {\bf r},\eta)\right\rangle = \frac{1}{2\pi^2}\int d\ln{k}  \frac{\sin{kr}}{kr} P(k),
\en
which is written in terms of the Heisenberg picture, and
\be
P(k,\eta) =  k^3 |f_{k}(\eta)|^2
\label{powerspectrum}
\en
is the power spectrum of $v$.

In Bohmian mechanics this quantity is obtained as follows.  First, let us denote $v(\eta,{\bf x};v_i)$, with $v_i$ a field on space, a solution to the guidance equations such that $v(\eta_i,{\bf x};v_i) = v_i({\bf x})$. If the initial field $v_i$ is distributed according to quantum equilibrium, i.e., $|\Psi(v_i,\eta_i)|^2$, then because of equivariance $v(\eta,{\bf x};v_i)$ will be distributed according to $|\Psi(v,\eta)|^2$. For such an equilibrium ensemble, we can consider the two-point correlation function
\begin{align}
&\left\langle v(\eta,{\bf x})v(\eta,{\bf x}+{\bf r})\right\rangle_{\rm B}  \\
& \quad = \int \mathcal{D} v_i |\Psi(v_i, \eta_i)|^2 v(\eta,{\bf x};v_i) v(\eta,{\bf x} + {\bf r};v_i) \\
& \quad = \int \mathcal{D} v |\Psi(v, \eta)|^2 v({\bf x}) v({\bf x} + {\bf r})
\end{align}
which  leads to the usual expression \eqref{twopoint} together with \eqref{powerspectrum} for the correlation function and the power spectrum of $v$, respectively.

The power spectrum determines the temperature fluctuations of the cosmic microwave background. Let us consider this in a bit more detail. Let $T({\bf n})$ denote the temperature of the cosmic microwave background in the direction ${\bf n}$, with ${\bar T}$ its average over the sky. The temperature anisotropy $\delta T({\bf n}) / {\bar T}$, where $\delta T({\bf n}) = T({\bf n}) - {\bar T}$, can be expanded in terms of spherical harmonics
\be
\frac{\delta T({\bf n})}{{\bar T}} = \sum^{\infty}_{l=2} \sum^{m=l}_{m=-l}
a_{lm} Y_{lm}({\bf n}) \,.
\en
The $a_{lm}$ are determined by the Mukhanov-Sasaki variable. The main quantity
used to study the temperature anisotropies is the angular power spectrum
\be
C^0_l = \frac{1}{2l+1} \sum_m |a_{lm}|^2 \,.
\en
In the standard treatments, one considers the operator ${\widehat C^0_l}$ and
compares the observed value for the angular power spectrum with $C_l = \langle \Psi | {\widehat C^0_l} |\Psi \rangle$, which is a function of the power spectrum \eqref{powerspectrum}. This is sometimes regarded as a puzzle, since the expectation value refers to an average over an ensemble of universes, while on the other hand we have only one sky to observe the microwave background radiation. It is sometimes claimed that this expectation value will agree with an average of the angular power spectrum seen for different observers at large spatial separations. While this may be true, it does not seem relevant, since we do not have observations from other places, just from earth.

What is relevant, however, is the variance. Since for the Bunch-Davies vacuum the modes are jointly Gaussian, then also the $a_{lm}$ are jointly Gaussian, and the variance of $C^0_l$ is given by \cite{lyth09} 
\be
(\Delta C^0_l)^2 = \frac{2}{2l+1}  C_l^2 \,.
\en
This means that one would expect the observed value to deviate from $C_l$ by an
amount of the order $\sqrt{2/(2l+1)}C_l$. We will hence have a greater uncertainty for
small $l$ (which corresponds to large angles over the sky, since the angle is
roughly $\pi/l $). For large $l$ the observed value must lie closer to the
expected value.

This kind of reasoning is justified in the Bohmian approach (while in the orthodox interpretation of quantum theory there remains a gap, viz., the measurement problem). Indeed, since in Bohmian mechanics the initial configuration $Q_0$ of the universe is a realization of (i.e., typical of) the $|\Psi|^2$ distribution, the $a_{lm}$ obtained from $Q_0$ are a realization of the joint distribution that follows from $|\Psi|^2$ which, as we assumed, is Gaussian in the case at hand. And for a realization of a Gaussian random variable, its deviation from the expectation value is indeed of the order of the root mean square.

\subsection{Quantum-to-classical transition in inflation theory}\label{qtoc}
In both inflationary models and bouncing models, the seeds of structure are the vacuum fluctuations (usually) described by the Bunch-Davies vacuum. During the evolution of the universe, these vacuum fluctuations become classical fluctuations. It is problematic to explain this transition within the context of orthodox quantum mechanics. Namely, the fluctuations are initially described by a wave function that is homogeneous and isotropic. The transition to classical fluctuations implies that these symmetries are somehow broken. However, since the Schr\"odinger dynamics preserves these symmetries, the only way this can happen in the context of orthodox quantum theory is through the collapse of the wave function. But when exactly does the wave function collapse in this case? This is the measurement problem, as was mentioned in the introduction. In the early universe there were no observers or measurement devices. Moreover, observers and measurement devices are supposed to originate from these vacuum fluctuations. Bohmian mechanics provides an elegant and simple solution to the problem \cite{hiley95,pinto-neto12a,pinto-neto14a}. The key is that in Bohmian mechanics there is more than the wave function. There are actual field fluctuations whose motion is determined by the wave function. Even though the wave function may be homogeneous and isotropic, the actual fluctuations generically are not. Moreover the field fluctuations start to behave classically during the expansion, as expected. We will explain this in the context of inflation theory \cite{hiley95,pinto-neto12a}. For bouncing models, see \cite{pinto-neto14a}.

According to the simplest inflationary models, the early universe has inflated almost exponentially driven by a single scalar field $\varphi$ (the inflaton field). The homogeneous and isotropic background is assumed classical (rather than described by Bohmian mechanics), as is usually done in inflation theory. The scalar perturbations are described by the Bunch-Davies vacuum which statisfies \eqref{xo2}. As discussed in the previous section, this state is completely determined by the function $f_k(\eta)$ which satisfies the classical mode equation \eqref{mode-f}, which depends on $z$. In many inflationary models (like power-law or slow-roll inflation), the behavior of $f_k(\eta)$ for physical wave lengths at early times, i.e., $\eta \rightarrow \eta_i$, is given by
\begin{equation}
\label{fka}
f_k(\eta) \sim \ee^{- \ii k \eta}\left(1 + \frac{A_k}{\eta} + \dots \right).
\end{equation}
As such, as follows from \eqref{soly}, the Bohmian modes are given by
\be
v_{\bf k}(\eta) \sim \left( 1+\frac{{\rm Re} A_k}{\eta} +  \dots \right)
\en
(in many inflationary scenarios, ${\rm Re} A_k = 0$ and the first order term disappears). So $v_{\bf k}$ tends to be time independent for $|\eta|\gg1$. (This is compatible with the fact that the Bohmian modes are stationary for the ground state of a quantized scalar field in Minkowski space-time~\cite{holland93b}.) Hence, the time dependence of the Bohmian modes is completely different from that of classical solutions, which oscillate for $|\eta|\gg 1$ and $k^2\gg z''/z$, see Eq.~(\ref{fka}).

The physical modes will grow larger during inflation and will eventually obtain wavelengths much bigger than the curvature scale, i.e., $k^2\ll z''/z$. When this happens, the behavior of $f_k(\eta)$ is approximately given by the so-called growing mode, i.e.,
\begin{equation}
\label{yqq}
f_k (\eta)  \sim A^g_k \eta ^{\alpha_g},
\end{equation}
where $\alpha_g <0$, so that $|f_k|$ equals $f_k$, up to a time-independent complex factor. As such, the Bohmian modes approximately evolve according to the classical mode equation~\eqref{mode-f}, so that the classical limit has been attained. The classical limit can also be investigated by examining the behavior of the quantum force and leads to the same result \cite{pinto-neto12a}.

\subsection{Observational aspects for matter bounces}\label{observational}
Matter bounces have been proposed as alternatives to inflation. First of all, they solve the horizon and flatness problems\footnote{The horizon problem refers to the question why the cosmic microwave background radiation had the same temperature in different regions of the primordial universe if the particle horizon of big bang models at that time was much smaller than the distances among these regions. The flatness problem concerns the question why the space-like hypersurface is so flat today if flatness is highly unstable under decelerated expansion. Both problems simply do not appear in bouncing models because particle horizons can be arbitrarily large in everlasting models and flatness is an attractor under decelerated contraction.} of the big bang scenario (which also motivated inflation). Second, just as in inflation, they yield a viable causal mechanism for the origin of the seeds of all structures in the universe and for the anisotropies of the cosmic microwave background radiation. The idea is that, in the far past, the matter content of the universe was homogeneously and isotropically diluted in an immensely large space which was slowly contracting. In this very soft matter contraction, space-time was almost flat and empty and the only source of inhomogeneities are quantum vacuum fluctuations. It has been shown that perturbations originating from such quantum vacuum fluctuations during a slow matter dominated contracting phase become almost scale invariant in the expanding phase.
This picture is called the matter bounce scenario. There are many proposals to cause the transition from contraction to the present expanding phase, most of them using another field (in addition to the regular matter field) which dominates at small scales and realizes the bounce. However, the presence of another field yields entropy perturbations, the relative fluctuations between the individual energy densities of the different fields, which are not usually treated correctly (for the correct treatment, see e.g.\ \cite{peter16}). Furthermore, in such classical bounce scenarios, primordial gravitational waves are also created, and they are as important as scalar perturbations, i.e., the ratio $r=T/S$ between the amplitude of primordial gravitational waves $T$ and the amplitude of scalar perturbations $S$ is approximately one. However, observations of the anisotropies of the cosmic microwave background indicate that the ratio $r$ is small, i.e., $r< 0.1$. This is a shortcoming of these classical bouncing models.

In section \ref{exponential}, we presented a quantum background model where, in the asymptotic past, the scalar field behaves like a matter fluid. Hence it is a matter bounce model, which leads to the observed almost scale invariant spectrum of scalar perturbations. Furthermore, this scalar field with exponential potential leads to a dark energy phase. As we have seen, the Bohmian quantum approach to this classical model opened a new possibility for bouncing scenarios with dark energy: the dark energy behavior may be a feature only of the expanding phase, where it can model the present observed accelerated expansion, and it was absent in the contracting phase. This is a very important property of this background because the presence of dark energy in the contracting phase of bouncing models makes problematic the imposition of vacuum initial conditions for cosmological perturbations in the far past of such models. For instance, if dark energy is a cosmological constant, all modes will eventually become bigger than the curvature scale in the far past and a vacuum prescription becomes quite contrived, possibly leading to divergences in the asymptotic past \cite{maier12}. In the above cosmological model, however, the presence of dark energy in the universe does not make problematic the usual initial conditions prescription for cosmological perturbations in bouncing models, the universe will always be dust dominated in the far past (running back in time, the dust repeller becomes an attractor) and vacuum initial conditions can be easily imposed in this era. Consequently, we get a well-posed problem to calculate the observed spectrum and amplitude of scalar cosmological perturbations in bouncing models with dark energy.

Let us now investigate what are the amplitudes of scalar perturbations and primordial gravitational waves in this model, and whether the classical problem with their ratio $r=T/S$ is overcome \cite{bacalhau17}. The primordial gravitational waves are described by the variable $\mu$ which satisfies similar equations as the Mukhanov-Sasaki variable $v$, see section \ref{bdvacuum}, with the scale factor $a$ playing the role of $z$. The Bunch-Davies vacuum state is also considered for these gravitational waves.

One can numerically calculate the modes of scalar perturbations and primordial gravitational waves up to the present epoch. The two-point correlation function \eqref{twopoint} for scalar perturbations is expressed in terms of $f_k$, which satisfies Eq.~\eqref{mode-f}. The analogous variable for gravitational waves, called $g_k$, satisfies
\begin{equation}
\label{mode-f2}
g_{\bf k}'' +  \left(k^2-\frac{a''}{a}\right) g_{\bf k}=0 .
\end{equation}
In order to understand qualitatively the final results, let us discuss what happens near the quantum bounce. When the universe contracts sufficiently and we get closer to the bounce in the contracting phase, the curvature of space-time increases a lot, and in this situation we arrive at the regime where $z''/z \gg k^2$ and $a''/a \gg  k^2$ (the so-called super-Hubble behavior). The solutions for the
scalar and tensor perturbations at leading order then read, respectively,
\begin{align}\label{solHa}
{\zeta}_k \equiv \frac{f_k}{z} &\approx A^1_k + A^2_k \frac{1}{R_H}\int \frac{\dd{}t}{x^2a^3}, \\
{h}_k \equiv \frac{g_k}{a} &\approx B^1_k + B^2_k \frac{1}{R_H}\int \frac{\dd{}t}{a^3},
\end{align}
where $x$ was defined in Eq.~\eqref{mud1}.

\begin{figure}[ht]
\centering
\includegraphics[width=0.8\textwidth]{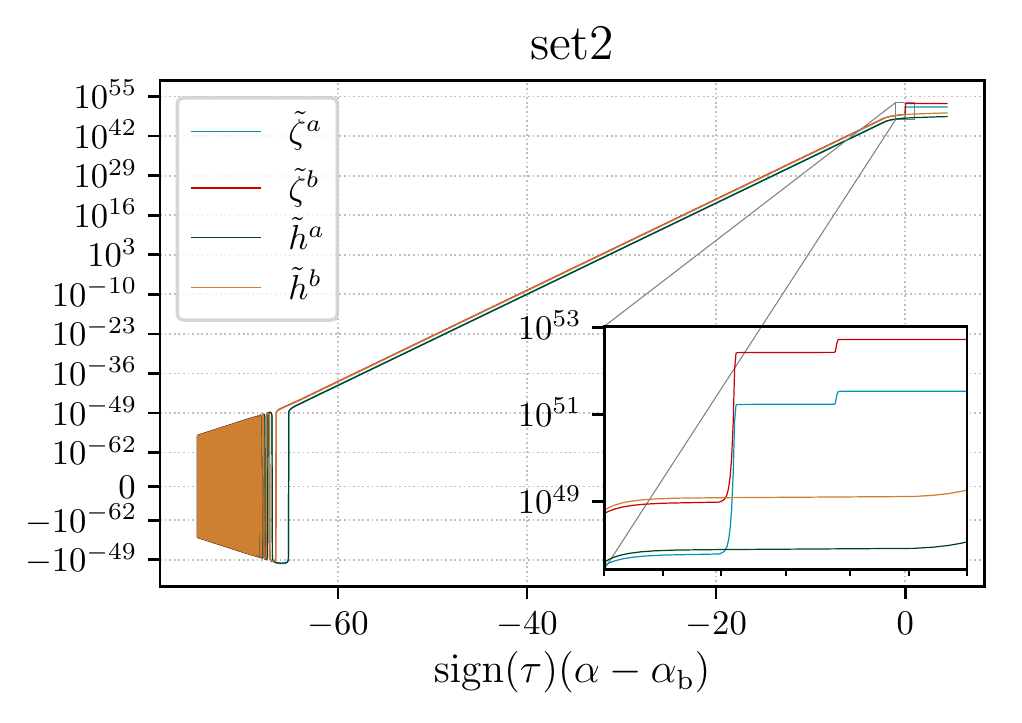}
\caption{Evolution of scalar and tensor perturbations in the background of case B. The indices 
$a$ and $b$ refer to the real and imaginary parts of the perturbation amplitudes. Scalar
and tensor perturbations grow almost at the same rate during classical contraction, but
at the quantum bounce the scalar perturbations are hugely enhanced over the tensor perturbations
due to the quantum effects (shown in the detail of the figure). After the bounce, the perturbations
get frozen. The final amplitudes of both perturbations are compatible with observations. }
\label{zeta_h}
\end{figure}

In the classical contracting branch of case B, $x$ varies between $1/\sqrt{2}$ and $1$, while the scale factor goes through a large contraction. In other words, the value of this integral will be dominated by the values of $a$ near the bounce phase, where $a$ attains its smallest value. Hence, for classical bounces the amplitude of scalar and tensor perturbations will be similar, and their ratio $r \approx 1$, contrary to the observation limit $r < 0.1$. However, in a quantum bounce, when the quantum phase begins, the value of $x$ is no longer restricted to $(1/\sqrt{2},\; 1)$, see Fig.~\ref{quantum-bounce2}, and it can increase the scalar perturbation amplitudes relative to the tensor perturbation amplitudes. Indeed,
during the classical matter phase, $1 / x^2 \approx 2$, and during the classical stiff matter
domination, $1/x^2 \approx 1$. Therefore, in the classical phase, the
presence of $1/x^2$ in the integral~\eqref{solHa} increases its value by
a maximum factor of two. On the other hand, throughout the quantum phase,
the $1/x$ factor can substantially increase the  scalar perturbations amplitudes relative to the tensor perturbations amplitudes, as can be
seen in Fig.~\ref{zeta_h}. The curve shows that the presence of
$1/x^2$ in the aforementioned integral will result in a sharp increase in
the scalar perturbation amplitude around $|\alpha-\alpha_b| \approx 10^{-1}$, where $\al_b$ is the value of the scale factor at the bounce.
This effect takes place close to the bounce.

This is a very important result: it shows that quantum cosmological effects may solve
problems which plague classical bouncing models, namely, large ratios of tensor to scalar
perturbations amplitudes. More than this, it shows that features of Bohmian trajectories
can lead to observational consequences and explain involved cosmological issues. Whether these conclusions can
also be reached under other quantum frameworks, is something yet to be verified.

The parameters of the theory can be adjusted in order to yield the right amplitudes and
spectral indices of scalar and tensor perturbations. Hence, a single scalar field with
a simple potential analyzed in the Bohmian framework yields a sensible
bouncing model with dark energy behavior and
correct perturbation amplitudes.

\section{Semi-classical gravity}\label{scg}
Semi-classical gravity is an approximation to quantum gravity where gravity is treated classically and matter quantum mechanically. In the usual approach to semi-classical gravity, matter is described by quantum field theory on curved space-time. For example, in the case the matter is described by a quantized scalar field, the state vector can be considered to be a functional $\Psi(\varphi)$ on the space of fields, which satisfies a particular Schr\"odinger equation
\be
\ii \pa_t \Psi(\varphi,t) = {\widehat H}(\varphi,g) \Psi(\varphi,t) \,,
\label{0.001}
\en
where the Hamiltonian operator ${\widehat H}$ depends on the classical space-time metric $g$. This metric satisfies Einstein's field equations
\be
G_{\mu \nu} (g) = 8\pi G \langle \Psi | {\widehat T}_{\mu \nu} (\varphi,g) |\Psi\rangle \,,
\label{0.002}
\en
where the source term is given by the expectation value of the energy-momentum tensor operator.

This will form a good approximation when the gravitational field and the matter field are behaving approximately classically, but will break down when the state of the matter enters in a macroscopic superposition. Namely, the Wheeler-DeWitt theory is linear, while the semi-classical approximation is not. So when matter is in a superposition of being in one location and another, then according to the Wheeler-DeWitt theory the state will be of the form $\Psi = (\psi_1(\varphi) \chi_1(g) + \psi_2(\varphi) \chi_1(g))/{\sqrt 2}$, where $\psi_i$ is the state of the matter at a particular location and $\chi_i$ the state of the corresponding gravitational field. However, according to the semi-classical theory the state will be of the form $\Psi = (\psi_1(\varphi)  + \psi_2(\varphi) )/{\sqrt 2}$, so that $\langle \Psi| {\widehat T}_{\mu \nu} |\Psi\rangle \approx \left( \langle \psi_1| {\widehat T}_{\mu \nu} |\psi_1\rangle + \langle \psi_2| {\widehat T}_{\mu \nu} |\psi_2\rangle \right)/2$ and hence the gravitational field is affected by two matter sources, one coming from each term in the superposition. As experimentally shown by Page and Geilker, the semi-classical theory becomes inadequate in such situations \cite{page81,kiefer04}.

In the context of Bohmian mechanics it is natural to consider another type of semi-classical approximation \cite{struyve15,struyve17a}. In this approximation, the back-reaction from the matter onto the gravitational field is by using the Bohmian energy-momentum tensor $T_{\mu \nu}(\varphi_B,g)$, where $\varphi_B$ is the actual Bohmian field configuration, in Einstein's field equations:
\be
G_{\mu \nu} (g) = 8\pi G T_{\mu \nu}(\varphi_{B},g) \,.
\label{0.003}
\en
While the resulting theory is still non-linear, it solves the problem with the macroscopic superposition. Namely, even though the matter wave function is in a macroscopic superposition of being at two locations, the Bohmian field $\varphi_{B}$ and the energy-momentum tensor $T_{\mu \nu}(\varphi_{B},g)$ will correspond to a matter configuration at one of the locations. This means that according to eq.\ \eqref{0.003} the gravitational field will correspond to that of matter localised at that location.

However, there is an immediate problem with this ansatz, namely that eq.\ \eqref{0.003} is not consistent. The Einstein tensor $G_{\mu \nu}$ is identically conserved, i.e., $\nabla^\mu G_{\mu \nu} \equiv 0$. So the Bohmian energy-momentum tensor $T_{\mu \nu}(\varphi_B,g)$ must be conserved as well. However, the equation of motion for the scalar field does not guarantee this. (Similarly, in the Bohmian approach to non-relativistic systems, the energy is generically not conserved.) The solution to the problem is that the usual expression for the Bohmian energy-momentum tensor is not the right source term in the Einstein field equations. The correct source term can in principle be derived by starting from the Bohmian Wheeler-DeWitt theory. However, in the derivation the gauge invariance, which in this case is the invariance under spatial diffeomorphisms (i.e., spatial coordinate transformations), should be dealt with, either by performing a gauge fixing or by working with gauge independent degrees of freedom. However, this is a notoriously difficult problem in the case of general relativity. In the case of mini-superspace model the spacial diffeomorphism invariance is elminiated and a consistent semi-classical approximation can be found straightforwardly.

A similar problem arises in the Bohmian semi-classical approach to scalar electrodynamics, which describes a scalar field interacting with an electromagnetic field. In this case, the wave equation for the scalar field is of the form
\be
\ii \pa_t \Psi(\varphi,t) = {\widehat H}(\varphi,A) \Psi(\varphi,t) \,,
\en
where $A$ is the vector potential. There is also a Bohmian scalar field $\varphi_{B}$ and a charge current $ j^\nu(\varphi_B,A)$ that could act as the source term in Maxwell's equations
\be
\pa_\mu F^{\mu \nu}(A) = j^\nu(\varphi_B,A)  \,,
\en
where $F^{\mu \nu}$ is the electromagnetic field tensor. In this case, we have $\pa_\nu \pa_\mu F^{\mu \nu} \equiv 0$ due to the anti-symmetry of $F^{\mu \nu}$. As such, the charge current must be conserved. However, the Bohmian equation of motion for the scalar field does not imply conservation. Hence, just as in the case of gravity, a consistency problem arises. In this case, the correct current can be derived by starting from the full scalar electrodynamics after eliminating the gauge freedom. For example, in the Coulomb gauge, there is an extra current $j^\nu_Q$ which appears in addition to the usual charge current and which depends on the quantum potential, so that Maxwell's equations read
\be
\pa_\mu F^{\mu \nu}(A) = j^\nu(\varphi_B,A) +  j^\nu_Q(\varphi_B,A) \,,
\en
which is consistent since the total current is conserved, because of the equation of motion for the scalar field.

Let us consider now the derivation of the semi-classical approximation of the mini-superspace model defined by \eqref{s5} and \eqref{s6}. For simplicity, we assume the gauge $N=1$. Given a set of trajectories $(\al(t),\varphi(t))$, the conditional wave function for the scalar field is $\chi(\varphi,t) = \psi (\varphi,\al(t))$. Using
\be
\pa_t \chi (\varphi,t) = \pa_\al \psi(\varphi,\al) \big|_{\al = \al(t)} \dot \al(t) \,,
\label{ms.11}
\en
and the Wheeler-DeWitt equation \eqref{s5}, we can write
\be
\ii \pa_t \chi =  {\widehat H}_M \chi + I \,,
\label{ms.12}
\en
where
\be
{\widehat H}_M = - \frac{1}{2\ee^{3\al}} \pa^2_\varphi + \ee^{3\al}V_M \,.
\label{ms.08}
\en
and $I$ a rest term \cite{struyve15}. When $I$ is negligible (up to a real time-dependent function times $\chi$), \eqref{ms.12} becomes the Schr\"odinger equation for a homogeneous matter field in an external FLRW metric. If furthermore the quantum potential $Q_G$ is negligible compared to other terms in eq.\ \eqref{s6.01}, we are led to the semi-classical theory:
\begin{align}
& \ii \pa_t \chi =  {\widehat H}_M \chi \,,\label{ms.14}\\
& \dot \varphi=   \frac{1}{\ee^{3\al}} \pa_\varphi S \,,\label{ms.15}\\
& \frac{1}{2} {\dot \al}^2 = \ka^2\left(\frac{1}{2} {\dot \varphi}^2 +V_M +Q_M\right) + V_G \equiv  - \frac{\ka^2}{\ee^{3\al}} \pa_t S  + V_G  \,. \label{ms.16}
\end{align}
This whole procedure is very similar to what was done in section \ref{quantumbackground}.

In the usual semi-classical approximation, one has \eqref{ms.14} and
\be
\frac{1}{2} {\dot \al}^2 = \frac{\ka^2}{ \ee^{3\al}}\langle \chi| {\widehat H}_M| \chi \rangle + V_G \,,
 \label{ms.17}
\en
with $\chi$ normalized to one (which is the analogue of \eqref{0.001} and \eqref{0.002} for mini-superspace). In \cite{struyve15}, an example was worked out showing that the Bohmian semi-classical approximation gives better results than the usual semi-classical approximation. As such the Bohmian semi-classical approximation may perhaps be used to find effects of quantum gravitational nature that are not present in the usual semi-classical approximation. Possible applications may be inflation theory (which will be discussed in the next section), which is described by a mini-superspace model with fluctuations. Usually the fluctuations are described quantum mechanically while the homogeneous background is described by a classical mini-superspace theory. Including the back-reaction from the fluctuations onto the homogeneous background will lead to corrections which may be testable. Such an investigation has been carried out in \cite{kiefer12}. A correction was found but is as yet unobservable. A Bohmian approach may perhaps improve on this result.

\section{Conclusion}
It is highly problematic to interpret canonical quantum gravity within the framework of orthodox quantum theory. The difficulty is that orthodox quantum theory merely makes predictions about outcomes of measurements and thereby relies on observers or measurement devices outside the quantum system of interest. When the quantum system of interest is the whole universe, there are no outside observers or measurement devices. As an alternative, there is Bohmian quantum gravity. Bohmian quantum gravity provides an objective description of the universe in terms of an actual space-time and matter fields (or particles), whose dynamics is determined by the universal wave function. This allows for a clear and unambiguous analysis of questions that are often rather ambiguous in the framework of orthodox quantum theory. Examples are the questions what the history of the universe is, whether it originated or will end up in a space-time singularity, how we can experimentally test the theory, etc.

In this chapter, we addressed various of these questions. We explained how Bohmian quantum gravity solves the problem of time and how it allows for the derivation of familiar time-dependent Schr\"odinger equations for subsystems in the universe.

We also analyzed the question of space-time singularities and described Bohmian solutions that are free of space-time singularities. This was done in the context of simplified models of quantum gravity which assume homogeneity and isotropy. In all the models we considered, bouncing solutions are possible which describe a contracting universe evolving into an expanding one. Such bouncing solutions may describe our actual universe.  We also considered deviations from homogeneity and isotropy by introducing linear perturbations. Using the Bohmian approach, we described how effective equations can be found describing the motion of these quantum perturbations in an external homogeneous and isotropic quantum background. The study of these perturbations is important since they leave an imprint on, for example, the cosmic microwave background radiation. This allows to distinguish between different theories. We presented the results of a recent study that investigated the perturbations for a background model with a matter bounce, where the bounce follows from the Bohmian dynamics. The results are in good agreement with observation and hence such a model may provide a serious alternative to inflation theory, which is currently the prevailing approach to early universe cosmology. It is important to stress that the Bohmian dynamics for the background is crucial to obtain these results.

The Bohmian approach also deals in a natural way with the classical limit. The classical limit is obtained whenever the space-time or matter degrees of freedom behave approximately classically. We applied this to the problem of the quantum-to-classical transition in inflation theory and bouncing theories. This problem is solved very simply in Bohmian mechanics and gives the usual results (which are problematic to justify within the context of orthodox quantum theory).

In the final section, we presented an approach to semi-classical gravity based on Bohmian mechanics. This approach goes beyond the usual semi-classical approach and might provide a new tool, unavailable in orthodox quantum theory, to probe quantum gravitational effects.

There are other applications of Bohmian quantum gravity that were not discussed here. One concerns the Boltzmann brain problem in cosmology. As was shown in \cite{goldstein15}, there is no such problem in the context of the Bohmian approach. Another application concerns the search for quantum non-equilibrium  \cite{valentini07,colin15,colin16}. While a typical initial configuration gives the usual quantum predictions described by the Born rule, other initial configurations may lead to non-equilibrium distributions which yield deviations from the Born rule. Since non-equilibrium distributions tend to evolve very quickly to equilibrium distributions \cite{valentini05,towler11}, one needs to find special systems where this is not the case. Such special systems might be found in cosmological context, in particular with the primordial perturbations for which non-equilibrium distributions may leave observational imprints on the cosmic microwave background. In some of these investigations, it was conjectured that some anomalies found in the observation of the anisotropies of the cosmic background radiation might be explained by these deviations. However, there is still a large debate on whether such anomalies are indeed statistically significant, or whether they are produced by other causes. So far, there is no experimental evidence for a violation of the Born rule.

\section{Acknowledgments}
NPN and WS respectively acknowledge the CNPq of Brazil and the Deutsche Forschungsgemeinschaft for financial support.

\end{document}